\documentclass[sigconf]{acmart}

\AtBeginEnvironment{quote}{\itshape}

\newcommand{\added}[1]{\textcolor{black}{#1}}

\usepackage{multirow}

\AtBeginDocument{%
  \providecommand\BibTeX{{%
    \normalfont B\kern-0.5em{\scshape i\kern-0.25em b}\kern-0.8em\TeX}}}

\copyrightyear{2024} 
\acmYear{2024} 
\setcopyright{rightsretained} 
\acmConference[ASSETS '24]{The 26th International ACM SIGACCESS Conference on Computers and Accessibility}{October 27--30, 2024}{St. John's, NL, Canada}
\acmBooktitle{The 26th International ACM SIGACCESS Conference on Computers and Accessibility (ASSETS '24), October 27--30, 2024, St. John's, NL, Canada}
\acmDOI{10.1145/3663548.3675641}
\acmISBN{979-8-4007-0677-6/24/10}

\acmSubmissionID{9392}

\begin{document}

\title[SMART-TBI]{SMART-TBI: Design and Evaluation of the Social Media Accessibility and Rehabilitation Toolkit for Users with Traumatic Brain Injury}


\author{Yaxin Hu*}
\orcid{0000-0003-4462-0140}
\affiliation{%
  \institution{Department of Computer Sciences\\University of Wisconsin--Madison}
  \streetaddress{Department of Computer Sciences, University of Wisconsin--Madison}
  \country{} 
}
\email{yaxin.hu@wisc.edu}

\author{Hajin Lim*}
\orcid{0000-0002-4746-2144}
\affiliation{%
  \institution{Department of Communication\\Seoul National University}
  \streetaddress{Department of Communication, Seoul National University, Seoul, South Korea}
  \country{} 
  }
\email{hajin@snu.ac.kr}

\author{Lisa Kakonge}
\orcid{0000-0003-4164-6716}
\affiliation{%
  \institution{School of Rehabilitation Science\\McMaster University}
  \streetaddress{Rehabilitation Science, McMaster University}
  \country{} 
}
\email{kakongel@mcmaster.ca}

\author{Jade T. Mitchell}
\orcid{}
\affiliation{%
  \institution{Dept. of Hearing \& Speech Sciences\\Vanderbilt University Medical Center}
  \streetaddress{Vanderbilt University Medical Center}
  \country{} 
}
\email{jade.t.mitchell@vanderbilt.edu}

\author{Hailey L. Johnson}
\orcid{0000-0003-1310-9948}
\affiliation{
  \institution{Department of Computer Sciences\\University of Wisconsin--Madison}
  \streetaddress{Department of Computer Sciences, University of Wisconsin--Madison}
  \country{} 
}
\email{hljohnson22@wisc.edu}
\author{Lyn S. Turkstra}
\orcid{0000-0002-6948-6921}
\affiliation{%
  \institution{School of Rehabilitation Science\\McMaster University}
  \streetaddress{Rehabilitation Science, McMaster University}
  \country{} 
}
\email{turkstrl@mcmaster.ca}

\author{Melissa C. Duff}
\orcid{0000-0003-1759-3634}
\affiliation{%
  \institution{Dept. of Hearing \& Speech Sciences\\Vanderbilt University Medical Center}
  \streetaddress{Vanderbilt University Medical Center}
  \country{} 
}
\email{melissa.c.duff@vanderbilt.edu}

\author{Catalina L. Toma}
\orcid{0000-0003-0714-312X}
\affiliation{%
  \institution{Department of Communication Arts\\University of Wisconsin--Madison}
  \streetaddress{Department of Communication Arts, University of Wisconsin-Madison}
  \country{} 
}
\email{ctoma@wisc.edu}

\author{Bilge Mutlu}
\orcid{0000-0002-9456-1495}
\affiliation{%
  \institution{Department of Computer Sciences\\University of Wisconsin--Madison}
  \streetaddress{Department of Computer Sciences, University of Wisconsin--Madison}
  \country{} 
}
\email{bilge@cs.wisc.edu}
\email{  }
\renewcommand{\shortauthors}{Hu and Lim, et al.}


\begin{abstract}
Traumatic brain injury (TBI) can cause a range of cognitive and communication challenges that negatively affect social participation in both face-to-face interactions and computer-mediated communication. In particular, individuals with TBI report barriers that limit access to participation on social media platforms. To improve access to and use of social media for users with TBI, we introduce the Social Media Accessibility and Rehabilitation Toolkit (\textbf{SMART-TBI}). The toolkit includes five aids (Writing Aid, Interpretation Aid, Filter Mode, Focus Mode, and Facebook Customization) designed to address the cognitive and communicative needs of individuals with TBI. We asked eight users with moderate-severe TBI and five TBI rehabilitation experts to evaluate each aid. Our findings revealed potential benefits of aids and areas for improvement, including the need for psychological safety, privacy control, and balancing business and accessibility needs; and overall mixed reactions among the participants to AI-based aids. 
\end{abstract}

\ccsdesc[500]{Social and professional topics~People with disabilities}
\ccsdesc[500]{Human-centered computing~Accessibility technologies}

\keywords{Traumatic Brain Injury (TBI), Accessibility, Social Media, Facebook}
 
\maketitle
\begin{figure}[!th]
  \centering
  \includegraphics[width=\columnwidth]{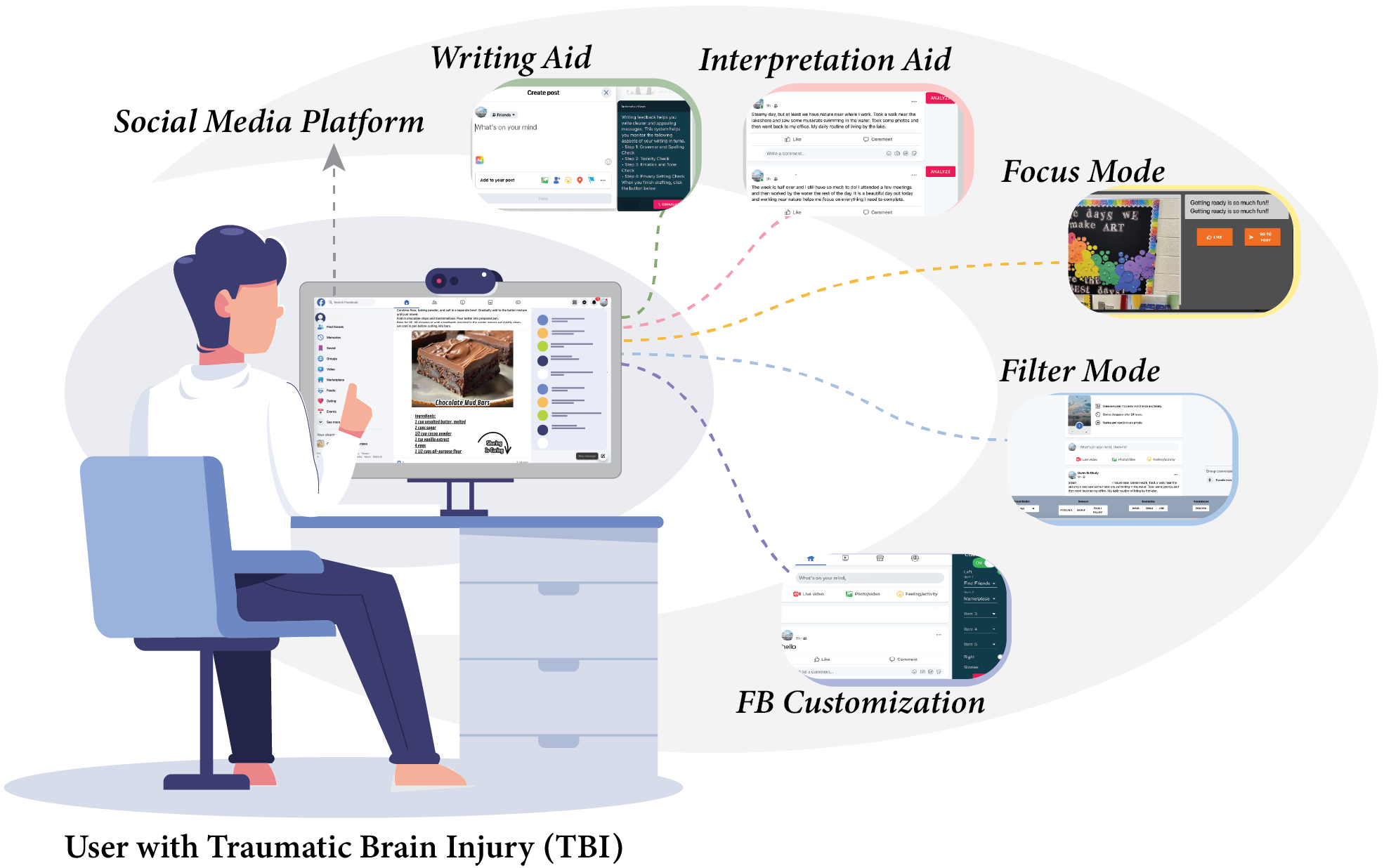}
  \caption{In this paper, we present the \textit{Social Media Accessibility and Rehabilitation Toolkit ({SMART-TBI})} that consists of five aids designed to serve as communication and cognitive support for individuals with TBI when using social media platforms. Eight users with TBI and five TBI rehabilitation experts evaluated our toolkit. The evaluation of these aids showed the usefulness of the aids as well as revealed usability challenges, informing our next steps in building accessible social media platforms for users with cognitive and communication challenges.}
  \Description{Figure depicts a drawing of an individual using a computer. Radiating from the computer screen are illustrations of the five aids in the SMART-TBI: The Writing Aid, The Interpretation Aid, Filter Mode, Focus Mode, and Facebook Customization}
  \label{fig:teaser}
\end{figure}

\section{Introduction}

Traumatic brain injury (TBI) is a significant public health concern, affecting approximately 69 million individuals every year worldwide \cite{dewan2018estimating,wongchareon2020impact}. TBI refers to damage caused to the brain as a result of an external force, and typically occurs through falls, car accidents, sports injuries, or assaults \cite{cdc2022get}. TBI can vary in severity from mild to severe, limiting an individual's functioning and leading to chronic cognitive, physical, and emotional impairments. Among these, cognitive and communication challenges are particularly debilitating, often interfering with an individual's ability to engage in everyday activities and social interactions \cite{macdonald2017introducing}.  

 Adults with TBI often report social isolation \cite{mukherjee2003women} and friendship loss \cite{salas2018relating} after injury. They may experience physical and cognitive limitations that make in-person social interactions challenging \cite{hoofien2001traumatic,turkstra2008measuring, hu2023investigating}. Thus, individuals with TBI could especially benefit from the social connection opportunities provided by computer-based communication technologies. Computer-mediated communication (CMC) is the use of social media, texting, or email to communicate with others, and is ubiquitous in today's society \cite{kaplan2010users}. Social media platforms have revolutionized how people establish social connections, collaborate, participate in social events, and obtain information in daily life \cite{ellison2015social,gil2012social,herring2002computer, hu2022polite}. Previous literature showed that social media use can enhance users’ social capital \cite{ellison2015social}, contribute to friendship maintenance \cite{sprecher2019social}, and stimulate social sharing \cite{choi2014social,choi2021understanding}, all of which could benefit adults with TBI. There is evidence that people with TBI use social media platforms such as Facebook and Twitter as frequently as those without a brain injury \cite{brunner2020if} and may even prefer these online interactions to face-to-face communication \cite{toma2024does}. Yet, the benefits of social media may not be fully accessible to these individuals due to their cognitive and communication impairments \cite{toma2024does}. These impairments can include cognitive overload, which makes processing information more challenging, and difficulties interpreting social cues, which are crucial for effective online communication \cite{brunner2015review,brunner2020if,tsaousides2011familiarity,feuston2017social,brunner2019kind}. \added{While prior research in this area has provided critical information about social media usage among individuals with TBI \cite{brunner2015review,brunner2019kind,brunner2021rehabilitation,morrow2021computer} and identified their challenges and needs \cite{brunner2022training,brunner2023developing, ahmadi2022facebook}, there is limited research on how to overcome those challenges and make social media accessible for individuals with TBI
\cite{brunner2015review, shpigelman2014facebook}.}

\added{To address this gap, we designed and built \textbf{SMART-TBI (Social Media Accessibility and Rehabilitation Toolkit for Traumatic Brain Injury)}, a suite of digital accessibility aids that aim to support adults with TBI-related cognitive and communication challenges so they can successfully use social media platforms. Our choice of accessibility aids was based on our prior collaborative research with adults with TBI, in which users envisioned social media accessibility supports \cite{ahmadi2022facebook, lim2023so, zhao2022designing}. SMART-TBI consists of five types of aids: Writing Aid, Interpretation Aid, Filter Mode, Focus Mode, and Facebook Customization. The toolkit was designed using Facebook because it was the most actively used social media platform among individuals with TBI \cite{morrow2021computer}, and can be easily integrated into a user's current social media practices.} 

We evaluated each aid in the toolkit with both users with TBI and rehabilitation experts specializing in TBI. During the user evaluation, we asked eight Facebook users with TBI to perform a series of tasks on the Facebook platform, both with and without using the aids. They also completed questionnaires to assess the usability and intention to use each aid and participated in interviews to provide feedback on potential improvements. Subsequently, we presented the SMART-TBI to five TBI experts and solicited their feedback on each aid with a questionnaire derived from the W3C Cognitive Accessibility Guidelines, which outline requirements and recommendations for making web content more accessible to people with cognitive and learning disabilities \cite{world2022all}. 

Our findings revealed the potential benefits of the toolkit in addressing diverse cognitive and communication challenges that individuals with TBI may encounter on social media platforms, while also indicating areas for improvement for each aid. In particular, the results highlighted SMART-TBI's potential to enhance social communications across various aspects, including self-presentation, organized use of social media, and distraction reduction. The results also shed light on design implications for future accessible social media design, emphasizing the need to promote psychological safety and privacy control, balance business profits with accessibility needs, and address mixed reactions to AI-based aids for toolkit adoption among individuals with diverse TBI needs.

Our contributions are as follows: 
\begin{itemize}
    \item \textbf{Toolkit design and development}: We designed and implemented the SMART-TBI that could be easily integrated into Facebook platforms. 
    \item \textbf{Insights for accessibility toolkit for individuals with TBI}: We evaluated the SMART-TBI with both users with TBI and TBI rehabilitation experts. Our findings highlighted both positive feedback and areas of improvement for each aid, offering design insights for the development and implementation of future accessible social media platforms for individuals with TBI.
\end{itemize}

\section{Related Work}

\subsection{Cognitive and Social Communication Challenges of  Individuals with TBI}

Individuals with TBI face a myriad of chronic cognitive, communication, and social cognitive challenges \cite{sohlberg2022transforming,togher2013cognitive,academy2020social}. Cognitive challenges from TBI may include difficulties in reasoning, attention and concentration, problem-solving skills, memory, and executive functions \cite{sohlberg2022transforming}. In particular, impairments in executive functions, such as inhibitory control (\textit{i.e.}, the ability to manage one's attention, thoughts, and behaviors to perform a necessary task) and working memory (\textit{i.e.}, holding information in mind and mentally manipulating it), can result in diminished focus and attention \cite{diamond2013executive}.  

In particular, cognitive-communication difficulties refer to challenges in communication related to language comprehension and production \cite{macdonald2017introducing}. Individuals with cognitive-communication challenges may struggle with speaking, word finding, understanding language, or expressing their thoughts effectively. For example, an individual with cognitive-communication difficulty might miss key details in written correspondences or repeat information \cite{dinnes2018writing}. These difficulties can lead to frustration and social isolation, making it harder for individuals to engage in meaningful interactions with others \cite{togher2023incog}.

Social communication, which relies on social cognition and language skills to engage in meaningful conversations across various social settings, is often impaired in individuals with TBI \cite{academy2020social,struchen2011examining,finch2016systematic}. Social cognition is crucial for interpreting social cues and communicating effectively \cite{sohlberg2022transforming,byom2012effects} involving recognizing emotions, predicting behaviors, and understanding others' intentions, encompassing components like the theory of mind and empathy \cite{byom2012effects}.

Additionally, many individuals with TBI struggle with behavioral self-regulation, including emotional modulation and impulse control \cite{mcdonald2021effect}. These difficulties together could lead to reduced social participation and lower life satisfaction \cite{dahlberg2006social}.

\subsection{Social Media Use and Individuals with TBI}

Individuals with traumatic brain injuries could benefit from social media platforms in mitigating social isolation. For instance, social media may lessen the cognitive, communication, and social demands of face-to-face interactions by providing more time to process information, formulate responses, and engage at their own pace without the immediate pressure of real-time conversation \cite{brunner2019kind}. Social media can also help individuals with TBI connect with others who have similar lived experiences and exchange social support \cite{morrow2021computer,brunner2019kind}.
Promisingly, prior research showed that individuals with TBI maintain social media accounts at similar rates as healthy individuals \cite{tsaousides2011familiarity,morrow2021computer} and are highly interested in using social media for various purposes, including social connection \cite{brunner2019kind,brunner2019content,morrow2021computer}. 

Nevertheless, individuals with TBI may also encounter various challenges when using social media due to their cognitive (\textit{e.g.}, attention, memory) and cognitive-communication (\textit{e.g.}, processing written information) impairments. One significant challenge is navigating the varied interfaces and features of social media platforms \cite{lim2023so}. The complexity of these interfaces can be overwhelming for individuals with TBI, and the abundance of information can be difficult to process \cite{brunner2019content}. Due to changes in cognitive function, individuals with TBI may experience difficulties in expressing themselves online, leading to reduced confidence in communication \cite{lim2023so,brunner2019content}. Additionally, individuals with TBI may experience a decreased ability to understand others' sentences or to read texts, which can further compound their social media challenges \cite{flynn2019characterizing, lim2023so}. For example, an individual with TBI with impaired social cognition may misinterpret a friend's sad post and comment with laughing emojis, failing to recognize the emotional context of the message \cite{clough2023emotion}. These challenges highlight the importance of providing sufficient resources to support individuals with TBI in navigating and taking advantage of social media platforms.

\subsection{Accessibility Support for Social Media}

Although Internet usage is common among individuals with TBI, there are still notable technological and access barriers in comparison to the general population \cite{baker2018internet}, limiting their ability to fully take advantage of the social benefits of social media platforms \cite{morrow2021computer}. Therefore, accessibility support for social media is crucial for individuals with TBI to access and utilize these platforms effectively. To improve social media accessibility for people with traumatic brain injuries, a recent study by \citet{lim2023so} adopted a participatory design approach to gather insights on technological tools that could improve the social media experience of users with TBI. Brunner and colleagues \citet{brunner2023developing} developed an online training course as part of the ``Social Brain Toolkit'' to support people with acquired brain injury to learn about social media use. Furthermore, \citet{zhao2022designing} proposed the design of four social media support aids to address challenges of sensory overload, memory loss, social communication, and lack of confidence in using social media faced by users with TBI. 

Research on individuals with cognitive and physical disabilities has revealed benefits and challenges of social media use similar to those experienced by people with TBI \cite{caton2016use, bassey2023perceptions, baumgartner2023if}. Common challenges include cognitive challenges, limited digital literacy, communication barriers, and the complexity of online interactions \cite{caton2016use, alfredsson2020access}. To address accessibility issues that stem from these challenges, researchers have developed solutions such as ``Endeavor Connect,'' a cognitively accessible Facebook interface designed for young adults with intellectual disabilities \cite{davies2015interface}.

As such, this work builds on a rich body of previous research on social media accessibility support. We designed and implemented SMART-TBI as web browser extensions that work on the web version of Facebook, providing users with essential support for social media use in their everyday lives. The SMART-TBI can be easily installed, offering a practical solution to the challenges faced by individuals with TBI when using social media platforms. While currently focused on Facebook, our approach has the potential to extend across various social media platforms.

\section{System Design - Accessibility Social Media Toolkit for TBI Users}

\begin{figure*}[!tb]
    \centering
    \includegraphics[width=\linewidth]{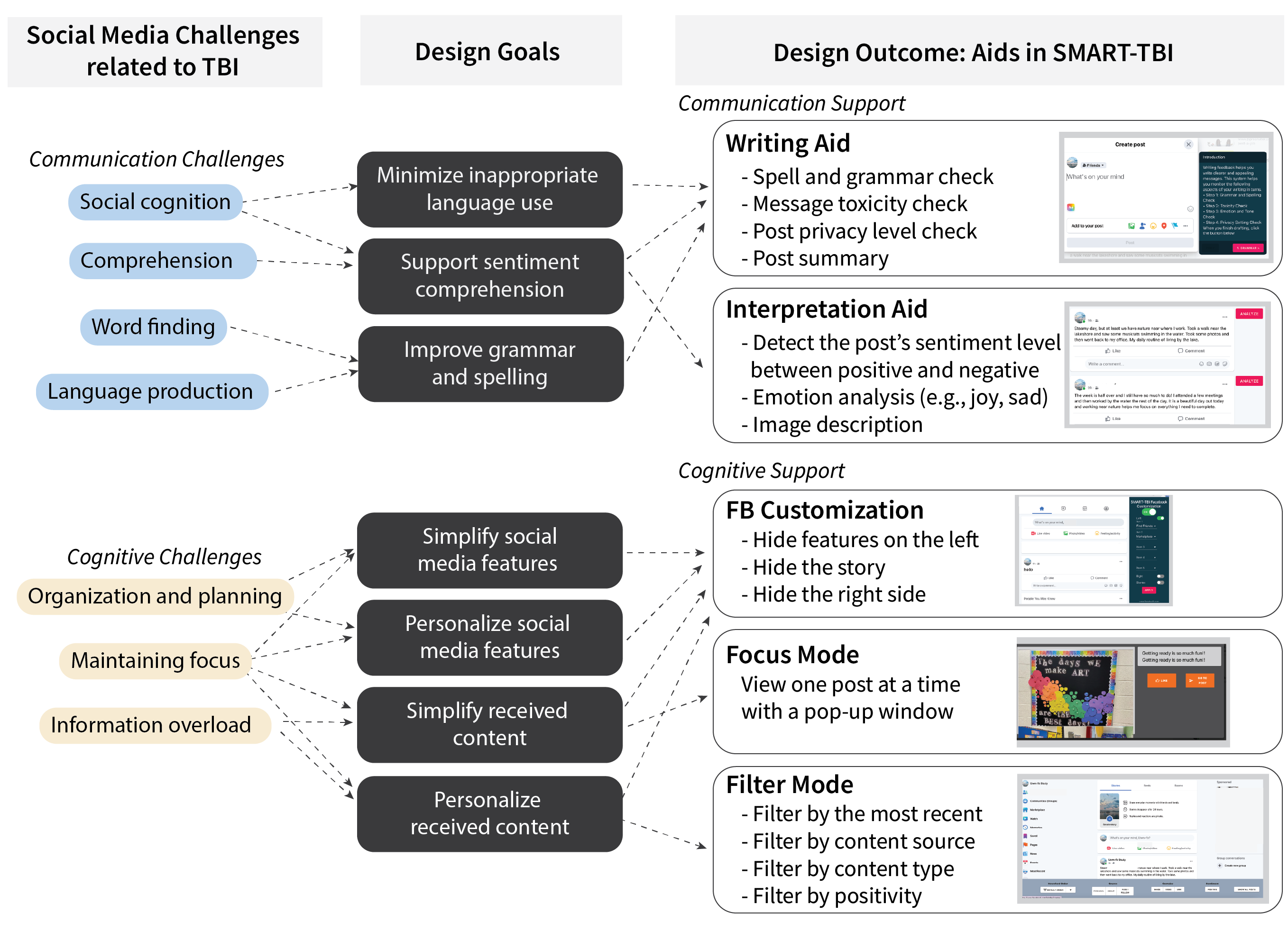}
    \caption{Design process for the SMART-TBI. Our designs were motivated by the social media challenges and needs by users with TBI identified in the prior work. Focusing on communication and cognitive challenges, we proposed design goals to overcome these challenges and generate the design of the SMART-TBI. \textbf{Left}: a series of challenges of social media use faced by individuals with TBI; \textbf{Middle}: design goals to overcome these accessibility challenges; \textbf{Right}: five aids to provide communication support and  cognitive support for social media use.}
    \Description{This figure illustrates the design process for the SMART-TBI, organized into three columns. The first column identifies social media challenges faced by individuals with Traumatic Brain Injury (TBI). The second column outlines the design goals developed to address these challenges. The third column presents five specific design aids that were created based on these goals, aiming to facilitate improved usability and engagement for users with TBI}
    \label{fig:process}
\end{figure*}

We designed and built a social media toolkit to meet the accessibility needs and challenges of TBI users that had been identified in prior literature \cite{lim2023so, zhao2022designing, brunner2023developing, davies2015interface}. We categorized the major challenges in social media use by individuals with TBI into two types: communication challenges and cognitive challenges. 

Communication challenges included challenges related to impairments in social cognition, language comprehension, and language production. These challenges might lead users with TBI to misinterpret the tone or intent of a written post and take sarcasm or humor literally, write messages that readers would consider inappropriate to the context, or overshare personal information. Cognitive challenges in using social media are mostly related to impairments in executive functions, leading to challenges in maintaining focus, planning, and managing information overload. As a result, users with TBI might struggle to organize their thoughts coherently in a post, leading to fragmented or confusing content. Similarly, challenges with planning could lead to impulsive posting without considering the consequences or the appropriateness of the content for a public platform. Information overload could have users with TBI become overwhelmed by details and irrelevant information surrounding posts---including text in sidebars---and give up on reading or posting content.

In addressing these challenges, we identified several design goals that we realized in five types of aids as shown in Figure~\ref{fig:process}. In particular, to address the communication challenges, we designed aids to help users comprehend social media content more accurately, enhance message construction, and minimize the creation of inappropriate content (\textit{e.g.}, offensive posts) \cite{lim2023so}. We identified three design goals: (1) minimizing inappropriate language use; (2) supporting sentiment comprehension; and (3) improving grammar and spelling. To address cognitive challenges, aids must assist users with TBI in managing information, facilitate navigation through social media features, and minimize distractions while viewing content \cite{lim2023so}. Accordingly, we established four design goals to address these cognitive challenges: (4) simplifying social media features; (5) personalizing social media features; (6) simplifying newsfeed content, and (7) and personalizing newsfeed content.

\begin{figure*}[!tb]
    \centering
    \includegraphics[width=\linewidth]{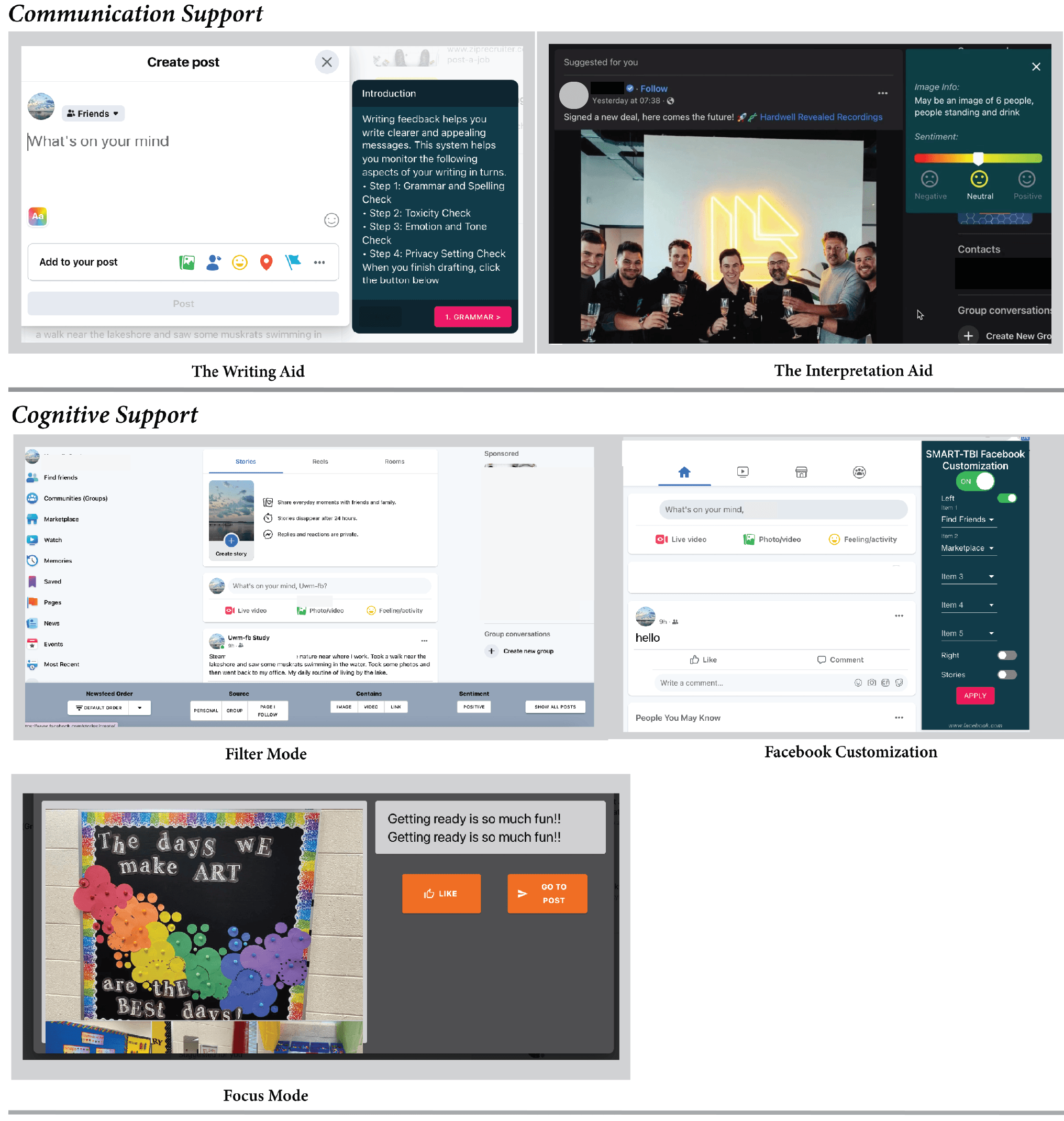}
    \vspace{5pt}
    \caption{An overview of the SMART-TBI. We developed two communication support aids and three cognitive support aids to assist the social media use for individuals with TBI. Communication support aids are \textit{The Writing Aid} and \textit{The Interpretation Aid}, and cognitive support aids are \textit{Filter Mode}, \textit{Focus Mode} and \textit{Facebook Customization}.}
    \Description{This visual overview depicts the SMART-TBI as it would appear on a social media platform. The toolkit comprises two types of support aids: communication and cognitive. For communication, the visuals of two aids are shown:  the Writing Aid and the Interpretation Aid, designed to enhance writing and interpretation skills respectively. For cognitive support, the visuals of three aids are included: Filter Mode, which screens distracting content; Focus Mode, which simplifies the user interface to enhance concentration; and Facebook Customization, which allows personalization of user experience on Facebook.}
    \vspace{15pt}
    \label{fig:five aids}
\end{figure*}

Based on these design goals, we developed five accessibility aids to assist individuals with TBI in using social media (Figure~\ref{fig:five aids}). Two aids were designed to provide communication support: (1) the \textit{Writing Aid}, which enabled users to perform four writing checks before posting on their Facebook feed, and (2) the \textit{Interpretation Aid}, which helped users interpret social cues (focusing on sentiment and emotion) within Facebook posts. Three aids were designed to provide cognitive support: (3) the \textit{Filter Mode}, which allowed users to customize their Facebook feed; (4) the \textit{Focus Mode}, which decluttered the Facebook news feed; and (5) the \textit{Facebook Customization}, designed to customize the Facebook layout. All five aids were implemented as Google Chrome extensions for use within the web version of Facebook. We detail the design and implementation of each aid in the following section. Demonstrations and implementation of each aid are provided in the GitHub repository\footnote{\url{https://smart-tbi.github.io/index.html} \\ \url{https://github.com/smart-tbi/smart-tbi}}. 

\newpage

\subsection{Communication Support}

\subsubsection{Writing Aid}

The goal of the Writing Aid was to help users with TBI compose postings that convey their intentions and meaning effectively and in socially appropriate ways by giving them feedback on various aspects of their post writing. In particular, the Writing Aid performed four types of writing checks, including (1) potential spelling or grammatical errors in their post; (2) potential toxicity within the language of the post; (3) the tone (\textit{e.g.}, positive, neutral, negative) and emotion type (\textit{e.g.}, happy, sad) of the post; and (4) the privacy settings of the post (\textit{e.g.}, public versus private). 

The detection of grammar errors, sentiment, and toxicity in the posts was achieved through external application programming interfaces (APIs) (\textit{i.e.}, Textgears API\footnote[1]{https://textgears.com/} for grammar check, IBM Watson NLP API\footnote[2]{https://www.ibm.com/products/natural-language-understanding} for sentiment and emotion detection, and perspective API for toxicity detection\footnote[3]{https://perspectiveapi.com/}). 

The Writing Aid interface starts to appear on the screen after users begin writing a post. Once they write a draft of their post, the aid guides users through the four writing checks, allowing them to recheck each step after any updates are made. After completing all four checks, the aid provides a full summary, including all writing-check results, followed by an opportunity for the user to review the changes to their post prior to posting the final draft. The accuracy of sentiment and emotion analysis was reported between 73\%--85\% \cite{abu2023emotion, carvalho2019off}, and the AUC-ROC scores \cite{bradley1997use} of the toxicity detection was reported between 0.97--0.99.\footnote[4]{https://developers.perspectiveapi.com/s/about-the-api-model-cards}
\added{While the spell-checking functionality in our Writing Aid powered by the Textgears API may not differ fundamentally from default browser spellcheckers, our goal was to provide a centralized, step-by-step approach to address the various considerations involved in writing posts. For individuals with TBI, navigating multiple aspects of writing, such as basic spell checking, toxicity and tone management, and privacy settings, can be overwhelming and prone to errors when distributed across different tools. Therefore, we consolidated these considerations into a single, guided process within the Writing Aid.}

\subsubsection{Interpretation Aid}

The Interpretation Aid was designed to help users understand the meanings and sentiments other users intend to convey in their Facebook posts. This aid used the same external API as the Writing Aid (IBM Watson NLP API\footnote[5]{https://www.ibm.com/products/natural-language-understanding}) to extract the emotion and sentiment of individual Facebook posts. Additionally, the alt texts of post images in the posts were extracted and shown to display the image details. 

While individuals with TBI are scrolling through their Facebook feed with this aid, an ``Analyze'' button appears alongside every post. If the user clicks the button, the Interpretation Aid interface appears beside the post that summarizes the sentiment analysis of the post, including the tone and emotion type. It also shows users the types of media used within the post (\textit{i.e.}, images, videos, links) and displays the alt text of images. 

\subsection{Cognitive Support}

\subsubsection{Filter Mode}
The Filter Mode aid was designed to help users customize their Facebook feed so that they only see preferred posts. The goal of this aid was to create a curated feed tailored to the user's interests, filtering out undesired or distressing content. 

Once users activate the Filter Mode, a gray options bar appears at the bottom of the Facebook screen. This bar contains four filtering options for users to choose from. The "Newsfeed Order" drop-down option allows users to view their Facebook newsfeed either in the default algorithmic order or chronologically based on the time of posting. The "Source" option lets users select the source of posts, such as personal posts, group posts, or page posts. The "Contains" option enables users to display posts that contain images, videos, or links. The ``Sentiment'' option allows users to filter their newsfeed to show only positive posts. Finally, there is an option to reset all previous filter choices.

\subsubsection{Focus Mode} \label{focus_mode_design}
The Focus Mode aid was designed to declutter the Facebook interface and help users limit their information intake by only focusing on the newsfeed. We had an initial design of the aid and we updated design based on the usability test and user feedback from the user evaluation. The initial design eliminated abstraction by showing only one post in the newsfeed at a time. When activated, it creates a screen overlay that includes a single post, an option to interact with the post via the ``Like'' button, and an option to view the next post in the user's feed. In the updated design, the user can see the full newsfeed list and the remainder of the Facebook interface is blurred in the background to minimize potential distractions.

\subsubsection{Facebook Customization}

The Facebook Customization aid was designed to streamline the visual interface, minimize clutter, and optimize the navigation experience of the social media platforms. It enabled users to toggle Facebook screen elements on and off, including elements on the left menu of the homepage (menu bar), the right section of the homepage (contact information), and the stories feed at the top of the website.

\section{System Evaluation} \label{systemeval}
To assess the SMART-TBI's potential usefulness and gather feedback for improvements, we conducted two system evaluation studies. In the first study, we recruited participants with TBI, and asked them to perform tasks on Facebook, with and without the aids from the SMART-TBI, and then provide feedback on each aid. For the second study, we recruited rehabilitation professionals who were experts in TBI, each of whom evaluated each aid. All study materials and protocols were administrated and approved by the University of Wisconsin-Madison Institutional Review Board (IRB). 

\begin{figure*}[!tb]
    \centering
    \includegraphics[width=\linewidth]{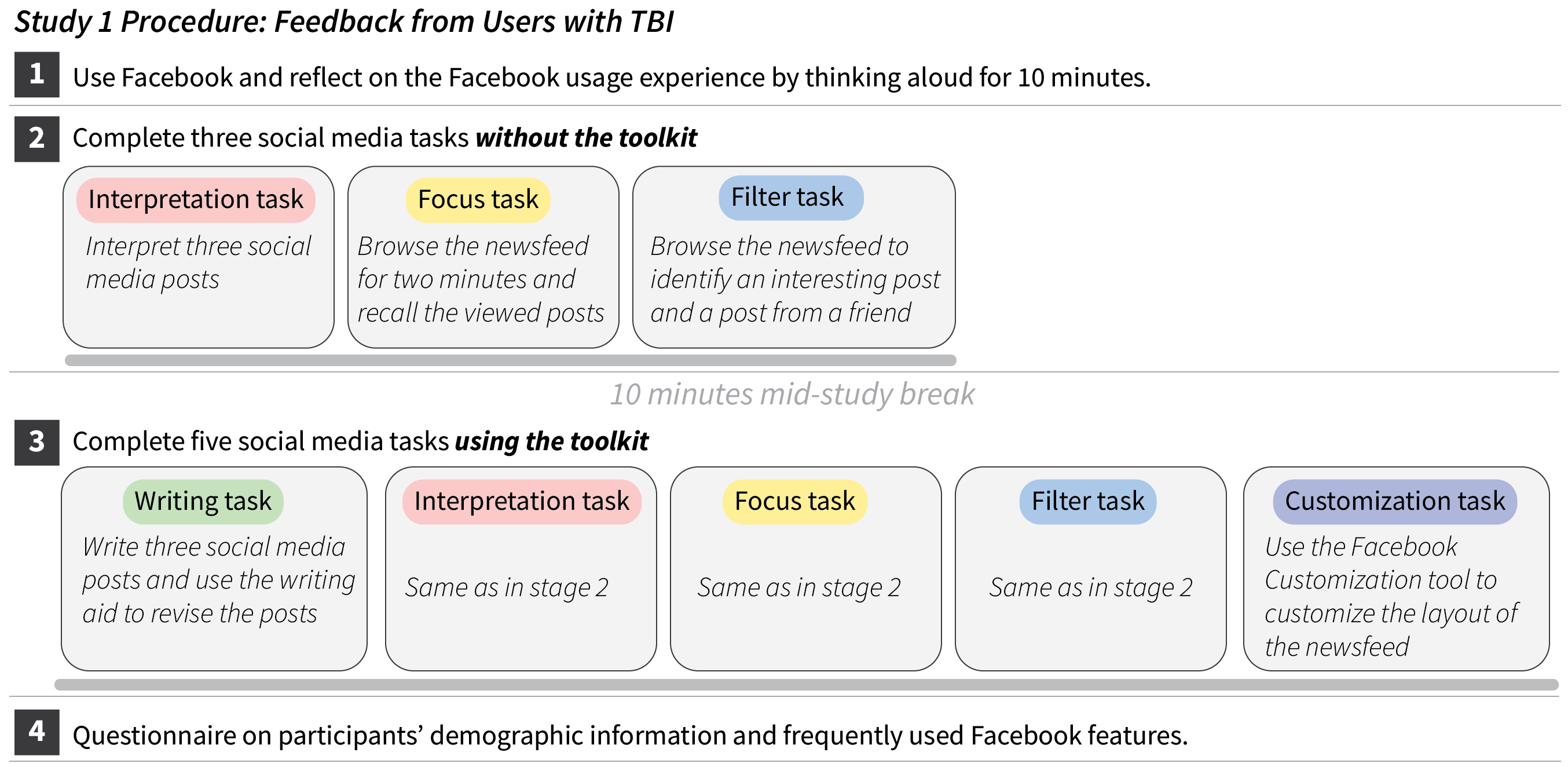}
    \caption{Procedure for Study 1: Feedback from users with TBI.}
    \Description{This figure displays the four stages of the formative study procedures for the SMART-TBI. Stage 1: Participants use Facebook and reflect on their experience. Stage 2: They complete three social media tasks without toolkit aids. Stage 3: They perform five social media tasks with toolkit aids. Stage 4: They fill out a demographic and Facebook usage questionnaire. }
    \label{fig:study1 procedure}
\end{figure*}

\subsection{Study 1: Feedback from users with TBI}

\subsubsection{Participants}
Participants were eight adults with moderate-to-severe TBI (3 women, 5 men; $M = 32.88$ years, $SD = 9.60$). All participants were from the continental US, native speakers of North American English, and recruited from a major hospital system registry \cite{duff2022value}. Inclusion criteria consisted of: (1) self-identification of English as a primary language; (2) no self-reported history of medical or neurological conditions, including brain diseases or premorbid language or learning disabilities affecting cognition; (3) possession of an active Facebook account; (4) knowledge of their Facebook log-in information; and (5) regular usage of Facebook. Exclusion criteria consisted of: (1) age under 18 years or over 55 years; and (2) an injury date less than six months from testing for participants with TBI. Participants older than 55 years were excluded to avoid the potential influence of cognitive changes and comorbid conditions associated with aging. Participants under 18 years were excluded to minimize cohort effects, as adolescents were likely to use other social media platforms. 

Medical records and intake interviews verified that the participants met the Mayo Classification System criteria for moderate-severe TBI \cite{malec2007mayo}. Barin injuries were classified as moderate-severe if at least one of the following criteria were met: (1) Glasgow Coma Scale (GCS) <13 within 24 hours of acute care admission; (2) positive neuroimaging findings (acute CT findings or lesions visible on a chronic MRI); (3) loss of consciousness (LOC) >30 minutes; or (4) post-traumatic amnesia PTA >24 hours. Participants with TBI were all in the chronic phase of injury (>6 months post-injury), and the average time post-injury was 68 months ($SD = 76.22$). Participant demographic details, injury characteristics, and information on the presence of long-term cognitive deficits are presented in Table~\ref{teb:demographics}. At the end of the study, each participant received \$20 USD as compensation for their participation.

\begin{table*}[!t]
    \caption{Demographic, injury, and Facebook usage information for participants with TBI.}
    \label{teb:demographics}
    \centering
    \footnotesize
    \renewcommand{\arraystretch}{2}
    \begin{tabular}{p{0.02\linewidth}p{0.02\linewidth}p{0.02\linewidth}p{0.075\linewidth}p{0.02\linewidth}p{0.095\linewidth}p{0.02\linewidth}p{0.02\linewidth}p{0.075\linewidth}p{0.4\linewidth}}
        \toprule
        \textbf{ID} & \textbf{Age} & \textbf{Edu} & \textbf{Etiology of TBI} & \textbf{TSO}  & \textbf{Race (Ethnicity)} & \textbf{Sex} & \textbf{Years on FB} & \textbf{FB usage pattern}  & \textbf{Cognitive and Communicative Challenges post-TBI}  \\ 
        \toprule
        P1 & 32 & 18 & Ped vs. auto & 68 & White (Not Hispanic) & F & 16 & Multiple times a week & Short-term or long-term memory loss; trouble concentrating or paying attention; Difficulty with language or speech production and thought processing; difficulty organizing or problem-solving; impulsiveness and lack of inhibition \\ 
        P2 & 54 & 16 & MVA & 227 & White (Not Hispanic) & M & 14 & Multiple times a day & Short-term or long-term memory loss \\ 
        P3 & 36 & 16 & Ped vs. auto & 130 & Black or African American (Not Hispanic) & M & 16 & Multiple times a day & Short-term or long-term memory loss; trouble concentrating or paying attention; difficulty with language or speech production and thought processing \\ 
        P4 & 29 & 18 & MVA & 61 & White (Not Hispanic) & F & 15 & Daily & Short-term or long-term memory loss; trouble concentrating or paying attention; difficulty with language or speech production and thought processing \\ 
        P5 & 28 & 12 & MVA & 20 & White (Hispanic) & M & 14 & Multiple times a day & Short-term or long-term memory loss; impaired judgment and perception; trouble concentrating or paying attention; difficulty with language or speech production and thought processing; difficulty organizing or problem-solving \\ 
        P6 & 26 & 12 & MVA & 13 & White (Not Hispanic) & M & 13 & Multiple times a day & Short-term or long-term memory loss; impaired judgment and perception; trouble concentrating or paying attention; difficulty with language or speech production and thought processing; difficulty organizing or problem-solving \\ 
        P7 & 35 & 12 & MVA & 12 & White (Not Hispanic) & M & 17 & Multiple times a day & None reported \\ 
        P8 & 23 & 12 & MVA & 13 & Black or African American (Not Hispanic) & F & 7 & Multiple times a day & Short-term or long-term memory loss; difficulty with language or speech production and thought processing \\ 
        \bottomrule
    \end{tabular}
    \caption*{\footnotesize{ID = participant ID number. Education (edu) reflects years of highest degree obtained. MVA = motor vehicle accident. MCC includes both motorcycle and snowmobile accidents. Ped vs. auto = participant was hit by a car while walking or running. Time since onset (TSO) is presented in months. F = female. M = male.}}
\end{table*}

\subsubsection{Study Procedure}

The study involving participants with TBI was conducted in a private lab space. Upon arrival, participants completed consent and payment forms in REDCap (Research Electronic Data Capture), a secure, web-based software platform designed to support data capture for research studies \cite{harris2009research}. The lab space was equipped with an HDR video camera, a participant laptop, and an experimenter laptop. The participant laptops were set up with Facebook open within a Chrome browser and a shared Google Doc for writing tasks. Each participant’s screen was shared in Zoom so that the experimenter and other research team members could monitor and record the session. After the consent process, participants were asked to sign into their personal Facebook account on the provided laptop. We then guided the following four stages in the study session (Figure~\ref{fig:study1 procedure}). 
\paragraph{\textbf{Stage 1} - Reflect on Facebook Usage}
In Stage 1, participants were asked to browse and use Facebook for ten minutes as usual without using aids to gather insights on their general Facebook experience. While using Facebook for ten minutes, an experimenter asked the participant questions to encourage reflection on their Facebook experience (\textit{e.g.}, ``Can you tell me your favorite part of Facebook''). 

\paragraph{\textbf{Stage 2} - Social Media Tasks without Using the Aids} \label{sec:task without the aid}
Within stage two, participants were asked to perform the following three tasks on Facebook without using any aids. 

\textit{Interpretation Task:} The first task involved interpreting three posts on the Facebook account, specifically created for this study (Appendix \ref{apd: interpretation task}). After participants viewed each post, we asked open-ended questions to participants, such as ``What emotions do you think of in this post?'' They were also asked to rate the sentiment of each post on a five-point Likert scale (\textit{i.e.}, Very Negative, Negative, Neutral, Positive, and Very Positive). Following completion of these tasks, participants rated their confidence in their judgments and ease in performing these interpretation tasks by answering the interpretation task questions in Table~\ref{tab: questionnaire for task feedback}.

\textit{Focus Task:} Participants were instructed to browse and scroll through their Facebook newsfeed for two minutes. Following that, the experimenter prompted them to recall and describe the posts they just viewed, using a few sentences (\textit{e.g.}, ``Can you recall the posts you just browsed''). Participants were then asked to provide feedback on the task by answering Focus task questions in Table~\ref{tab: questionnaire for task feedback}.

\textit{Filter Task:} In the third task, participants were asked to browse their Facebook newsfeed and inform the experimenter whenever they found an interesting post. Then, they were asked to refresh the page again and find a post from a friend \footnote{Based on our observations, Facebook always prioritizes unseen content at the top. Thus, we asked participants to refresh the newsfeed page before and between tasks.}. Following this, participants were asked to provide feedback on the task by answering Filter task questions in Table~\ref{tab: questionnaire for task feedback}.

After completing the three tasks described above, the participants were asked to identify their top five most frequently used menus in the left panel of the Facebook web browser. The total duration of stage 2 ranged from 20 to 30 minutes for each participant. 

\paragraph{\textbf{Stage 3} - Social Media Tasks Using the SMART-TBI aids} \label{sec:task with the aid}
During stage 3, participants first repeated the same Interpretation Task, Focus Task, and Filter Task as in stage 2 using the aids: Interpretation Mode, Focus Mode, and Filter Mode, respectively. Before starting each task, the experimenter \added{introduced the aid that would be used for the task and demonstrated how to use it step by step. Then the user had a trial session using the aid until they felt comfortable and familiar with the tool.} After the participant finished the trial session, the experimenter instructed the participant to proceed with the tasks. Additionally, participants completed a Writing Task where the experimenter asked the participant to write three hypothetical posts on provided topics (Appendix \ref{apd: writing task}). Additionally, they were asked to use the Writing Aid to improve their writing. After writing each post, the participant was asked a series of questions about their experience in the writing task (see Table~\ref{tab: user task feedback}).  Lastly, participants worked on the Facebook Customization Task that asked them to customize the layout of their Facebook main page using the Facebook Customization aid. 

After each task, participants evaluated each aid with the System Usability Scale (SUS) \cite{brooke1996sus} and answered open-ended questions (\textit{e.g.}, ``How was your experience with the tool?'') to reflect on their impressions on each aid. After completing all five tasks in stage 3, the participants were instructed to rank the five aids in order of most helpful to least helpful. The total duration of stage 3 ranged from 45 to 60 minutes. 

\paragraph{\textbf{Stage 4} - Questionnaires: Demographic Information and Facebook Usage}
Stage 4 consisted of participants answering demographic questions (\textit{e.g.}, gender, age, race, education, and employment status) and questions related to their TBI (\textit{e.g.}, time since injury, cause of injury, and challenges since injury). This stage ended with surveying participants about their usual Facebook usage (\textit{e.g.}, usage amount, reasons for usage, changes to social media usage after their TBI). The study session for each participant lasted between 90 minutes to two hours. 

\begin{table*}[!t]
    \caption{Demographic Information of Traumatic Brain Injury (TBI) Expert Participants}
    \label{tab:experts}
    \centering
    \renewcommand{\arraystretch}{1.2}
    \small
    \begin{tabular}{p{0.05\linewidth}p{0.1\linewidth}p{0.075\linewidth}p{0.1\linewidth}p{0.1\linewidth}p{0.1\linewidth}p{0.3\linewidth}}
    \toprule
        \textbf{Expert ID} & \textbf{TBI Experience} & \textbf{Number of Years} & \textbf{Age} & \textbf{Description} & \textbf{Race} & \textbf{Education} \\ \toprule
        EP1 & Daily & 30 & 55-64 years old & Female & White or Caucasian & Graduate or professional degree (MA, MS, MBA, PhD, JD, MD, DDS, etc.) \\ 
        EP2 & Daily & 12 & 35-44 years old & Female & White or Caucasian & Graduate or professional degree (MA, MS, MBA, PhD, JD, MD, DDS, etc.) \\ 
        EP3 & 4-6 times a week & 8  & 25-34 years old & Female & Asian & Graduate or professional degree (MA, MS, MBA, PhD, JD, MD, DDS, etc.) \\ 
        EP4 & Once a week & 5 & 35-44 years old & Female & White or Caucasian & Graduate or professional degree (MA, MS, MBA, PhD, JD, MD, DDS, etc.) \\ 
        EP5 & Once a week & 3  & 25-34 years old & Female & White or Caucasian & Graduate or professional degree (MA, MS, MBA, PhD, JD, MD, DDS, etc.) \\ \bottomrule
    \end{tabular}
    \caption*{\footnotesize{\textit{Expert ID} = ID number assigned to the expert. \textit{TBI Experience} = How often do you interact with people with TBI or develop/design technologies for people with TBI or other individuals with cognitive challenges? \textit{Number of Years} = How many years have you worked with people with TBI? \textit{Age} = How old are you? \textit{Description} = How do you describe yourself? - Selected Choice. \textit{Race} = Choose one or more races that you consider yourself to be. \textit{Education} = What is the highest level of education you have completed?}}
    \vspace{-12pt}
\end{table*}

\subsection{Study 2: Feedback from TBI Experts}
\added{The SMART-TBI was envisioned not only as a standalone tool but also as a potential asset in therapy settings. We foresee its use in training people with TBI to utilize social media platforms effectively by TBI experts. Consequently, we sought TBI experts' views on how the SMART-TBI might impact current therapy practices in communication and cognitive rehabilitation support.}

\subsubsection{Recruiting TBI Experts}
We recruited a convenience sample of rehabilitation professionals with expertise in brain injury, all known to the research team members. The rehabilitation professionals ($n = 5$) were all speech-language pathologists who provided rehabilitation support to either a pediatric or adult population of individuals with TBI, within community or rehabilitation center settings. The rehabilitation professionals had an average clinical experience of $M = 11.6$ years ($SD = 10.83$), ranging from 3 to 30 years (See Table~\ref{tab:experts}). 

\subsubsection{Study Procedure}
In the expert evaluation study, TBI rehabilitation professionals were invited to join a remote session via a Zoom link provided by the research team. A brief description of the purpose of this study was followed by instructions on the study procedures and the recording of sessions. After signing the consent form, we present participants with the W3C's Cognitive Accessibility Guidelines \cite{world2022all} and ask them to evaluate the current web version of Facebook based on the guideline. Under each objective in the guideline, there is a list of the design requirements with checkboxes. For example, the objective five ``Help users focus'' has four design requirements: (1) ``Limit interruptions''; (2) ``Make short critical paths''; (3) ``Avoid too much content''; and (4) ``Provide information so a user can complete and prepare for a task.'' Participants checked out the boxes of the design requirement items if they thought Facebook fulfilled the corresponding design requirement. 

Next, we asked participants to share their web browser screens and helped them install the five aids on their Chrome browsers. We then introduced each aid and demonstrated how to use it. Following that, participants used the aid to complete a task the aid was designed for. For example, they used ``Focus Mode'' to browse the news feed and ``Writing Aid'' to compose a post. Following this, participants evaluated each aid with the W3C's Cognitive Accessibility Guidelines \cite{world2022all}. After evaluating each aid, participants filled out the questions regarding their experiences with TBI and provided demographic information. The study procedures lasted between 30 minutes to 1 hour, and participants received \$50 USD in the form of gift cards as compensation.

\subsection{Data Analysis}
We employed a mixed-methods approach to collect and analyze five distinct sets of data described in Section~\ref{systemeval}: (1) interviews with TBI users from Study 1; (2) task performance data for TBI users from Study 1; (3) survey responses from TBI users from Study 1; (4) interviews with TBI experts from Study 2; and (5) survey responses from TBI experts from Study 2. 

\subsubsection{Quantitative Analysis} \label{sec: quant analysis}

\begin{table}[]
\small
\caption{Questionnaire for user's social media task feedback}
\begin{tabular}{l}
\toprule
\textbf{Writing task}                                       \\
1. The message is very clear.                          \\
2. The message says what I mean to say.                \\
3. The message will be well-received by others.        \\
4. The message will receive many likes and comments.   \\
5. The message will not offend other people.           \\
6. Others will understand this message.                \\
7. This message is well-written.                       \\
                                                       \\
\midrule
\textbf{Interpretation task}                                \\
1. I am confident in my interpretation of these posts. \\
2. What the writers intend to say is clear to me.      \\
3. These posts were easy to understand.                \\
                                                       \\
\midrule
\textbf{Filter task \& Focus task }                       \\
1. The task I just did is simple.                      \\
2. The task I just did is mentally demanding.          \\
3. I feel there is time pressure.                      \\
4. I did well on the task.                             \\
5. I feel frustrated with the task.                     \\
\bottomrule
\end{tabular}\
\label{tab: questionnaire for task feedback}
\vspace{-12pt}
\end{table}

\paragraph{User Evaluation: Aid Ranking and SUS} 
Participants were asked to rank each aid from one to five and evaluate each aid with the System Usability Scale (SUS). To accommodate the cognitive challenges associated with participants' TBI, we employed a simplified three-point Likert scale (Disagree, Neutral, and Agree) for each statement in the SUS. 

\paragraph{User Evaluation: Tasks Feedback and Task Performance Measures} \label{sec: user eval task feedback and performance}
We developed questionnaires with a three-point Likert scale (1--3; 1 = Disagree, 2 = Neutral, 3 = Agree) for each aid (Table \ref{tab: questionnaire for task feedback}). For the writing task, we utilized a seven-item questionnaire and derived two scales that measured the perceived quality of the written post (items 1, 7; Cronbach's $\alpha = 0.62$) and how well the message is received by other people (items 3, 5; Cronbach's $\alpha = 0.61$). For the Interpretation task, we used a three-item questionnaire to evaluate the confidence level, the clarity of the post, and the ease of the task. For the Focus task and Filter task, we developed a five-item questionnaire and derived two scales that measured the perceived ease of the task (items 1, 2, 5; Cronbach's $\alpha = 0.86$) and success of the task (items 3, 4; Cronbach's $\alpha = 0.94$). We hypothesized that the user would perceive the writing task to have higher quality using the Writing aid than without the aid; the user would have a higher level of confidence in the Interpretation task using the Interpretation aid than without the aid; the user would perceive the task to be easier and more successful using the Focus Mode and Filter Mode than without the aids. 

We also collected and analyzed task performance data for the Focus Mode and Filter Mode aids, and we evaluated their effectiveness pre- and post-use of the aids. Specifically, we collected the number of posts recalled by the participant after viewing the newsfeed for two minutes before and after applying Focus Mode to the newsfeed. We hypothesized that the user would recall more posts using the Focus Mode than without the aid. We also counted the time the user found an interesting post and the time to locate a post from their friend before and after they applied the Filter Mode to customize the feed. We hypothesized that the user would spend less time locating the posts using the Filter Mode than without the aid.  

\paragraph{Expert Evaluation: W3C Survey Result}
W3C's Cognitive Accessibility Guidelines contain eight objectives and each objective has a list of design requirements necessary to meet the objective. We customized the W3C questionnaires for each aid and removed irrelevant items. For example, ``Objective 8: Support adaptation and personalization.'' is not applied to the Writing Aid and the Interpretation Aid because these two aids were designed to provide communication support rather than improve the personalization of social media use. Therefore, Objective 8 was removed when participants evaluated these two aids. 

In analyzing the expert questionnaires for the guidelines for Facebook in general, as well as for the five aids, we calculated the percentage of design patterns met for each objective. For example, to determine if the ``Focus mode'' met the objective ``Help users focus,'' we scored it as 50\% when one participant answered ``Yes'' to two design patterns out of four. We also transcribed and open-coded the comments that expert participants provided while using each of our five aids. 

\subsubsection{Qualitative Analysis}

\paragraph{Interviews with Users with TBI} \label{sec: user analysis}

We recorded the full study sessions with users with TBI. Audio files were first transcribed with an automatic transcription tool\footnote[6]{https://otter.ai}, and then one researcher from the team verified the transcriptions and corrected errors. This researcher further segmented the transcriptions according to the study procedures, differentiating between responses related to questionnaire items and answers given during the experimenter's interview questions.

We analyzed the transcriptions using thematic analysis \cite{clarke2021thematic, braun2006using}. Two coders first independently open-coded three data samples (more than 10\% of the data) at the sentence level and then merged their codes to develop the initial codebook. Any disagreements during this phase were resolved through discussion. The same coders continued to process the remaining data individually, updating the codebook as new codes emerged. The final codebook included categories detailing participants' social media usage patterns, the challenges they faced using social media due to TBI, and their feedback on each aid. Given we have specific goals to understand the usefulness and usability for each aid in the toolkit, we follow the deductive approach \cite{clarke2021thematic, braun2006using} to generate themes focusing on particular aspects, \textit{i.e.}, the perceived usefulness, usability challenges, and suggested new functions for each aid. 

\paragraph{Interviews with TBI experts} 

We had both online and in-person study sessions with TBI expert participants. The online study session was hosted through teleconferencing technology (Zoom), and we recorded the full study sessions. In one in-person study session, the experimenter experienced technical issues with audio recorders, and the experimenter took field notes for the participant's response. The interview data were transcribed using an automatic transcription tool, and one researcher verified the accuracy of the transcripts. We followed the same approach of deductive thematic analysis as the analysis for interviews with TBI users as described above.

\section{Results}
This section presents the findings from our evaluation of the SMART-TBI involving eight users with moderate-severe TBI and five TBI rehabilitation experts. Both qualitative and quantitative results highlighted the strengths of each aid and the areas that require improvements.

\subsection{Quantitative Results} \label{sec: quant results}
\begin{table}[]
\caption{User Ranking for Each Aid}
\small
\label{tab: aid ranking by user}
\begin{tabular}{lrrrrrrrr}
\toprule
                 & \multicolumn{1}{l}{P1} & \multicolumn{1}{l}{P2} & \multicolumn{1}{l}{P3} & \multicolumn{1}{l}{P4} & \multicolumn{1}{l}{P5} & \multicolumn{1}{l}{P6} & \multicolumn{1}{l}{P7} & \multicolumn{1}{l}{P8} \\
 \toprule
Writing          & 4                      & 1                      & 1                      & 1                      & 1                      & 3                      & 4                      & 4                      \\
\midrule
Interpretation   & 5                      & 4                      & 3                      & 2                      & 3                      & 2                      & 3                      & 3                      \\
\midrule
Focus            & 1                      & 3                      & 5                      & 5                      & 4                      & 5                      & 5                      & 1                      \\
\midrule
Filter           & 3                      & 2                      & 4                      & 3                      & 5                      & 4                      & 2                      & 2                      \\
\midrule
Facebook Customization & 2                      & 5                      & 2                      & 4                      & 2                      & 1                      & 1                      & 5        \\
\bottomrule
\end{tabular}
\end{table}

\subsubsection{User Evaluation: Aid Ranking and SUS}

Our participants in Study 1 (users with TBI) ranked the five aids according to their preference after the study session (Table~\ref{tab: aid ranking by user}). On average, the Writing Aid ranked the highest, followed by Facebook Customization in second place. The Interpretation Aid and Filter Mode had the same ranking score in third place, while Focus Mode ranked the lowest. The preference towards the aids was also reflected in the SUS score reported in Appendix~\ref{apd: SUS scores}. In response to the statement, ``I think that I would like to use the system frequently,'' 50\% of participants agreed to the Writing Aid and 83.3\% for Facebook Customization. In contrast, only 25\% of the participants selected ``Agree'' for the Interpretation Aid, Focus Mode, and Filter Mode. These evaluations pointed to the usability challenges faced by participants and are reported in detail in \S\ref{sec:user feedback}.

\begin{table*}[!h]
\caption{Statistics for User Social Media Task Feedback. Participants provided ratings from 1--3 (1 = Disagree, 2 = Neutral, 3 = Agree) in answering the questions from Table~\ref{tab: questionnaire for task feedback}.}
\footnotesize
\label{tab: user task feedback}
\begin{tabular}{p{0.15\linewidth}p{0.2\linewidth}p{0.05\linewidth}p{0.05\linewidth}p{0.05\linewidth}p{0.05\linewidth}p{0.05\linewidth}p{0.05\linewidth}p{0.05\linewidth}p{0.05\linewidth}}
\toprule
\multicolumn{1}{c}{\multirow{2}{*}{Aid}} & \multicolumn{1}{c}{\multirow{2}{*}{Scale}}          & \multicolumn{2}{c}{Before using the aid}        & \multicolumn{2}{c}{After using the Aid}            & \multicolumn{3}{c}{Statistics}                                           \\
\multicolumn{1}{c}{}                     & \multicolumn{1}{c}{}                                & \multicolumn{1}{c}{M} & \multicolumn{1}{c}{Std} & \multicolumn{1}{c}{M}    & \multicolumn{1}{c}{Std} & \multicolumn{1}{c}{t}  & \multicolumn{1}{c}{DF} & \multicolumn{1}{c}{p}  \\
\midrule
\multirow{2}{*}{Writing Task}              & How the message will be received by others           & 2.79                  & 0.39                    & 2.77                     & 0.51                    & 0.44                   & 23                     & 0.67                   \\
                                         & Post is well-written                                & 2.85                  & 0.27                    & 2.85                     & 0.38                    & 0                      & 23                     & 0.5                    \\
\midrule
\multirow{3}{*}{Interpretation Task}      & I am confident in my interpretation of these posts. & 2.72                  & 0.70                    & 3.00                     & 0.00                    & -1                     & 6                      & 0.18                   \\
                                         & What the writers intend to say is clear to me.      & 2.71                  & 0.70                    & 2.71                     & 0.70                    & \multicolumn{1}{l}{NA} & 6                      & \multicolumn{1}{l}{NA} \\
                                         & These posts were easy to understand.                & 2.71                  & 0.70                    & 2.86                     & 0.35                    & -1                     & 6                      & 0.18                   \\
\midrule
\multirow{2}{*}{Focus Task}              & Ease of the task                                    & 2.96                  & 0.11                    & 2.71                     & 0.42                    & 2.645                  & 7                      & 0.02*                   \\
                                         & Performance of the task                             & 2.75                  & 0.66                    & 2.94                     & 0.17                    & -1                     & 7                      & 0.82                   \\
\midrule
\multirow{2}{*}{Filter Task}             & Ease of the task                                    & 2.96                  & 0.11                    & \multicolumn{1}{l}{2,54} & 0.8                     & 1.33                   & 7                      & 0.11                   \\
                                         & Performance of the task                             & 2.69                  & 0.66                    & 2.69                     & 0.66                    & 0                      & 7                      & 0.5                  
\\
\bottomrule
\end{tabular}
\end{table*}
\subsubsection{User Evaluation: Task Feedback and Task Performance}\label{sec-user-eval}

One-tailed paired samples t-tests were used for the evaluation of task feedback and task performance by TBI participants. In terms of the task feedback, participants perceived that Focus Task were significantly easier without the aid than with the aid [$t(7) = 2.645, p = 0.02$], suggesting the potential usability challenge of the Focus Mode. We did not find statistically significant differences in participants' feedback on the Interpretation and Filter tasks before and after using the aid. The results are presented in Table~\ref{tab: user task feedback}. 

In addition, we did not find statistically significant differences between participants' task performance with and without using the aid, including the number of the posts viewed within two minutes before ($M=1.4$, $STD=1.1$) and after ($M=0.9$, $STD=0.9$) using the Focus Mode [$t(7) = 2.65, p = 0.98$]; the time spent to find an interesting post before ($M=21.75\ seconds$, $STD=16.6\ seconds$) and after ($M=27.4\ seconds$, $STD=16.3\ seconds$) using the Filter Mode [$t(7) = -0.55, p = 0.70$]; Time spent to find a post from a friend before ($M=24.6\ seconds$, $STD=30.4\ seconds$) and after ($M=27.8\ seconds$, $STD=22.0\ seconds$) using the Filter Mode [$t(7) = -0.18, p = 0.57$]. Therefore, all hypotheses in \S\ref{sec: quant analysis} were not supported. We report participants' feedback for each aid in more detail, including the perceived usefulness and usability challenges in \S\ref{sec:user feedback} to inform the areas of improvement and design implications for the accessibility toolkit.

\subsubsection{Expert Evaluation: W3C Survey Result}

\begin{table*}[!tb]
    \caption{W3C evaluation results from TBI experts study}
    \label{tab:w3c experts result}
    \centering
    \footnotesize
    \renewcommand{\arraystretch}{2}
\begin{tabular}
{p{0.15\linewidth}p{0.07\linewidth}p{0.07\linewidth}p{0.07\linewidth}p{0.07\linewidth}p{0.07\linewidth}p{0.07\linewidth}p{0.07\linewidth}p{0.07\linewidth}p{0.07\linewidth}} 
    \toprule
\textbf{Aid Name}                       & \textbf{Objective 1} & \textbf{Objective 2} & \textbf{Objective 3} & \textbf{Objective 4} & \textbf{Objective 5} & \textbf{Objective 6} & \textbf{Objective 7 }& \textbf{Objective 8} & \textbf{Overall}\\
\toprule
Facebook               & 20\%        & 20\%        & 60\%        & 28\%        & 25\%        & 20\%        & 20\%        & 15\%    & 26\%    \\
Writing Aid            & 77\%        & 80\%        & 72\%        & 64\%        & 70\%        & 80\%        & 60\%        & NA      & 71.9\%    \\
Interpretation Aid     & 74\%        & 80\%        & 80\%        & 46\%        & 60\%        & 20\%        & NA          & NA     & 60\%     \\
Focus Mode             & 66\%        & 80\%        & 28\%        & 26\%        & 100\%       & 60\%        & 10\%        & 67\%    & 60\%    \\
Filter Mode            & 57\%        & 60\%        & 88\%        & 48\%        & 70\%        & 80\%        & 20\%        & 73\%    & 67.2\%    \\
Facebook Customization & 74\%        & 65\%        & 68\%        & 56\%        & 100\%       & 80\%        & 10\%        & 93\%  & 73.8\%     \\
\bottomrule
\end{tabular}
\caption*{\footnotesize{\textbf{Objective 1}: Help Users Understand What Things are and How to Use Them; \textbf{Objective 2}: Help Users Find What They Need; \textbf{Objective 3}: Use Clear and Understandable Content; \textbf{Objective 4}: Help Users Avoid Mistakes and Know How to Correct Them; \textbf{Objective 5}: Help Users Focus; \textbf{Objective 6}: Ensure Processes Do Not Rely on Memory; \textbf{Objective 7}: Provide Help and Support; \textbf{Objective 8}: Support Adaptation and Personalization}}
\end{table*}

The results for expert evaluation of Facebook and each aid based on the W3C Cognitive Accessibility Guidelines are presented in Table~\ref{tab:w3c experts result}. Overall, the experts' evaluation of Facebook reported low scores, with only 26\% of the requirements being fulfilled among all the objectives. Experts evaluated each aid's function and design and reported relatively high scores for the Writing Aid and Facebook Customization. Specifically, more than 70\% of the criteria under the overall objectives for Writing Aid (71.9\%) and Facebook Customization (73.8\%) were met. The Interpretation Aid, Focus Mode, and Filter Mode fulfilled 60.0\%, 60.0\%, and 67.2\% of the overall objectives, respectively. 

\begin{table*}[htp]
    \caption{Qualitative Findings: User feedback and expert feedback on each aid}
    \label{tab: all qual findings}
    \centering
    \footnotesize
    \renewcommand{\arraystretch}{1.2}
    \small
    \begin{tabular}{p{0.075\linewidth}p{0.12\linewidth}p{0.22\linewidth}p{0.25\linewidth}p{0.25\linewidth}}
    \toprule 
\textbf{Aid Name}                    & \textbf{Would like to use the aid frequently (User)}                                                        & \textbf{Perceived Usefulness}                                                                                                                                                                                                                                                                                                                              & \textbf{Usability Challenges and Suggestions}                                                                                                                                                                                                                                       & \textbf{Suggested new functions}                                                                                                                                                                                                                   \\
\midrule
Writing Aid            & \begin{tabular}[t]{p{\linewidth}}Agree (50\%)\\ Neutral (25\%)\\ Disagree (25\%)\end{tabular}     & \begin{tabular}[t]{p{\linewidth}}Helped with message construction and spell checks (P6, 8; EP2)\\ Support sentiment check (P5, 7; EP1, 4)\\ Support self-presentation (P1, 5)\\ Provide different interpretation perspective (P6; EP3, 5)\\ Privacy check (P7)\\ Learning grammar (P2) \\ Summarize the post (P7)\end{tabular} & \begin{tabular}[t]{p{\linewidth}}Be able to go back and redo the checks after modification (P2, 7, 8; EP3, 4)\\ Further support the content correction (P2; EP1, 3) \\ Need learning support (P2; EP2) \\ Mis-detection (P3, 4, 7; EP4) \\ Improve the instruction and wording (EP1--3) \\ The font is small (EP1)  

\end{tabular}                             

& \begin{tabular}[t]{p{\linewidth}}Automatically correct or rephrase the messages (P3, 5, 8)\\ Provide alternative word recommendations (P2, 5)\\ 

Privacy control and prevent oversharing (EP2, 5) \\
Support clinical practice (EP3)\\
Use speech input (EP5)

\end{tabular}                                                                 \\
\midrule
Interpretation Aid & \begin{tabular}[t]{p{\linewidth}}Agree (25\%)\\ Neutral (0\%)\\ Disagree (75\%)\end{tabular}        & \begin{tabular}[t]{p{\linewidth}}Support comprehension (P2, 6, 8)\\ Provide different interpretation perspective (P6; EP5) \\  Simplified the content (EP5) \\
Support social communication (EP1, 4)
\end{tabular}                                                                                                                                                                             & \begin{tabular}[t]{p{\linewidth}}Inaccurate interpretation results (P2, 3, 7, 8; EP1, 2, 5)\\ Improve image description (P2; EP3) \\ Clarify the content being analyzed (EP1, 3, 4) 
\end{tabular}                                                                                                                   & \begin{tabular}[t]{p{\linewidth}}Explain the reason behind the interpretation (P1, 5)\\ Support more content types (P2, 8, 5) \\
Report hate speech (EP3) 
\end{tabular} \\
\midrule 
Focus Mode             & \begin{tabular}[t]{p{\linewidth}}Agree (25\%)\\ Neutral (0\%)\\ Disagree (75\%)\end{tabular}     & \begin{tabular}[t]{p{\linewidth}}Easier to distinguish ads and posts (P8)\\ Help to read in more details (P8) 
\\Help to focus (EP1, 4, 5) 
\\Simplify the page (EP2, 3)
\end{tabular} 

& \begin{tabular}[t]{p{\linewidth}}Need learning support (P1; EP3) \\ 
Increase font size and photo size (P5, 8)\\ Post navigation challenge (P3, 4, 5, 8) 
\end{tabular} &    \begin{tabular}[t]{p{\linewidth}}                           Apply to other pages (EP5) \\
Privacy control and prevent oversharing (EP5)               \end{tabular}                                                                                                                                                              \\
\midrule
Filter Mode            & \begin{tabular}[t]{p{\linewidth}}Agree (25\%)\\ Neutral (12.5\%)\\ Disagree (62.5\%)\end{tabular} 

& \begin{tabular}[t]{p{\linewidth}}Help to filter out ads (P8)\\ Help with organized ways of using the social media (P8; EP3) \\
Ensure psychological safety (EP3, 5) \\ Narrowing down content (EP3) \\ Sorting is helpful (EP4, 5)
\end{tabular}                                                                                                                                                                                                            & \begin{tabular}[t]{p{\linewidth}}
Lost track of seen posts (P8; EP2)\\ 
Did not filter properly (P3, 4, 7; EP4, 5) \\
Improve labels (EP1, 2) \\ The font is small (EP3) \\ Add user instructions (EP2) 

\end{tabular}                                                                                         & Provide additional filtering options (P2, 5, 8)                                                                                                                                                                                                  \\
\midrule 
Facebook Customization & \begin{tabular}[t]{p{\linewidth}}Agree (83.3\%)\\ Neutral (0\%)\\ Disagree (16.6\%)\end{tabular}  & \begin{tabular}[t]{p{\linewidth}}Customize the newsfeed (P3, 7, 8; EP1, 3, 4, 5)\\ Make the user more focused (P5; EP1) 
\end{tabular}                                                                                                 & \begin{tabular}[t]{p{\linewidth}}
Need learning support (P7) \\ Improve labels (EP1, 2) 

\end{tabular}                                                                                                                      & \begin{tabular}[t]{p{\linewidth}}Provide verbal interaction (P4)\\ Filter out more specific posts: certain groups, more left side options (P5, 7) \\ Clean up ads (EP5) \end{tabular}       \\                                                                  
\bottomrule
\end{tabular}
\end{table*}
\subsection{Feedback for Each Aid in the SMART-TBI} \label{sec:user feedback}

Based on the usability challenges indicated in \S\ref{sec: quant results}, participants further provided feedback on how to address these challenges and improve each aid in the SMART-TBI. This section highlights the findings from the qualitative feedback followed by key design insights from TBI users and experts respectively. For each aid, we first present participants' overall attitudes, drawing on their responses to the interview question, \textit{``How do you like the aid?''}, and their ratings from the SUS statement, \textit{``I think I would like to use the system frequently.''} This is followed by detailed comments from participants about their experiences focusing on the perceived usefulness of the aid, usability challenges, and suggested new features. The findings are summarized in Table~\ref{tab: all qual findings}.

\subsubsection{\textbf{Feedback for Writing Aid}}
\paragraph{User Feedback} During the study, participants used the Writing Aid for spelling and grammar checks for the writing tasks. 
All participants found the aid useful, and four out of eight participants agreed that they would like to use the aid frequently (P1, 5, 7, 8). The aid was reported to support message construction and spell checks (P6, P8), facilitate sentiment analysis (P5, P7), and enhance privacy settings (P7). One participant noted it provided \textit{``a different perspective on how others might view it''} (P6), thus indicating that the design objectives of the Writing Aid could be attained to some degree. Additionally, two participants emphasized the need for self-presentation in using social media (P1, P5) and thought that the aid could \textit{``help you change your story''} (P5) and confirm that \textit{``Is this character (myself) polished enough to be on someone's network''} (P1).

Meanwhile, TBI user participants identified two major usability challenges: inaccurate grammar and sentiment analysis of the content (P3, P4, P7) and the inconvenience of repeating writing checks after modifying posts (P2, P7, P8). For example, P4 experienced inaccurate grammar suggestions that incorrectly flagged slang she wrote in her posts, such as \textit{``gonna''} and \textit{``Ima}.'' P7 suggested a potential design improvement, proposing a single button to restart the sequence of checks after making edits. He stated, \textit{``I would add a button at the end to go back to the start...in case you want to redo something and maybe recheck a specific section before you post it.''} (P7). Additionally, P2 desired more learning support, stating that, \textit{``it was frustrating at first''} (P2), and wanted the aid to provide more \textit{``specific''} instructions for improving the content.   

\paragraph{Expert Feedback}
Four out of five expert participants (EP1--3, EP5) provided positive feedback on the overall functionality of the aid, highlighting its capability to reduce communication errors and enhance writing clarity (EP1, EP2) and provide useful perspectives to tweak the written post (EP3, EP5). Further, EP5 emphasized the need to use the writing aid to support social communication; she commented: \textit{``Some of the big issues is kind of that impulsivity and not being able to kind of check and correct their own errors and stuff when posting...First of all, does it make sense, what it's saying, but also kind of how that might come across to other people, as well}.'' (EP5).

Experts (EP1--5) also pointed out usability challenges of the aid based on the W3C guideline and suggested improvements, focusing on TBI users’ cognitive and sensory needs. These suggested improvements included increasing font sizes (EP1); using simple and clear sentences for instructions (EP1--3); defining keywords used in the aid (\textit{e.g.}, \textit{``toxicity''}) (EP3); and providing more detailed suggestions or automatically implement the suggestions after the grammar checks (EP1, 3). EP2 suggested rewording the aid's instruction in a \textit{``more direct or simple''} way because of the diverse literacy levels of the patients she had worked with. EP2 shared, \textit{``A lot of my patients that I work with who use social media a lot of the time, they come from backgrounds. Some of them don't have a lot of education.''} (EP2). 

Furthermore, experts (EP2, EP5) pointed out that oversharing personal information often happens for TBI users and suggested that the Writing Aid could \textit{``add in that extra level of security''} (EP5) to ensure the user's safety, such as sending out alerts when oversharing activity was detected (EP5). EP5 also suggested using verbal input to overcome the challenge of word-finding.

\subsubsection{\textbf{Feedback for Interpretation Aid}}
\paragraph{User Feedback}The Interpretation Aid was found to support comprehension of posts and images (P2, P6, P8), particularly for longer messages (P2), and two participants agreed that they would like to use it frequently (P1, P8). For example, P8 reported that \textit{``Right after the accident, I had a really hard time understanding what people meant with their words...This [aid] would have been really comforting to me.''} P6 appreciated that the aid provided a different perspective of the post, stating, \textit{``it's nice to see how other people, same as earlier, like, could interpret what you're saying.''} 

The major usability challenge reported by participants for the Interpretation Aid was disagreement with the results of sentiment analysis for the content (P2, P3, P7, P8). For example, P2 disagreed with the classification of the emotion type \textit{``joy''} for the post in the first interpretation task. P8 thought a post to be \textit{``positive''} while the aid predicted it to be \textit{``negative}.'' Additionally, P2 noted confusion about the image information and thought the description of its emotion should be shortened. 

Notably, two participants (P1, 5) suggested that the aid provide more detail and explain the reason behind the interpretation results. For example, P1 commented: \textit{``I want them to emphasize more, like, you know, why are they negative, why are they positive?''} (P1).

\paragraph{Expert Feedback}
Three expert participants (EP1, EP4, EP5) reacted positively to the Interpretation Aid. They appreciated that the aid could support social-emotional communication (EP1, EP4) and summarize the posts for individuals with TBI (EP5). For example, EP1 described how the aid can help users react to other people's social media posts, stating, \textit{``The aid would help them to know what the post is sort of about and whether to `like' it or not. Maybe even determine whether it’s a like or a `love'. Like if they were to learn that the greener it is the more chance the person is looking for a `love', versus a yellow one the person is looking for a 'like'''} (EP1). Similar to the TBI users, experts also identified usability challenges for the aid, such as the need to improve the accuracy of interpretation results (EP1, EP2, EP5) and clarify the analyzed content (EP1, EP4).

\subsubsection{\textbf{Feedback for Focus Mode}}
We evaluated both versions of the Focus Mode as described in \S\ref{focus_mode_design}. The initial design was evaluated by our TBI user participants and the updated design was evaluated by our TBI expert participants.  

\paragraph{User Feedback} Three participants (P1, P3, P8) reacted positively towards the Focus Mode aid, and two participants reported that they would like to use it frequently (P1, P8). For example, P8 described the aid \textit{``amazingly helpful''} (P8) and shared, \textit{``I read more captions than I would have if I was scrolling the newsfeed. They're bigger and easier to see.''} 

On the other hand, five participants (P2, P4--7) held negative views towards the aid, finding it \textit{``unnecessary''} (P2) and \textit{``cumbersome''} (P5). Two major usability challenges reported by multiple participants were the post navigation for viewing the full content (P3--5, P8) and the unresponsiveness of the buttons (P2, P6--8). P5 described how the buttons for post navigation were inaccessible for users with motor impairments, stating \textit{``If you have, you know, in nimble fingers, and you can't really do that, or if you're old and don't know how to do it''} (P5). In addition to the above challenges, two participants (P5, P8) requested to increase the size of the post creator's profile photo and name texts.

\paragraph{Expert Feedback}
All TBI expert participants provided positive feedback for the Focus Mode, highlighting its ability to enhance the usability of social media by streamlining the individual’s feed (EP1--5) and the simplicity of use (EP1). For example, EP4 commented: \textit{``Wow. Okay, this is the best one. Wow, I can really focus on the posts.''} (EP4). Similarly, EP5 mentioned that  \textit{``You can literally just focus on scrolling through your newsfeed. Yeah. This is cool.''} (EP5). The major usability challenge reported by TBI experts participants (EP3, EP4) is the interactivity of the blurred area. EP4 described how she accidentally clicked in the blurred area and went to an unexpected page: \textit{``The blur, I accidentally clicked and they went to a sponsor page.''} (EP4). 

\subsubsection{\textbf{Feedback for Filter Mode}}
\paragraph{User Feedback} 
Four participants (P1, P4, P6, P8) held positive views towards the aid,  two participants (P2) held neutral views, and three (P3, P5, P7) reacted negatively to the aid. Two participants (P1, P8) agreed that they would like to use the aid frequently. In particular, P8 found it helpful that advertisements and posts were distinguished after applying the aid, stating that \textit{``I think the advertisements being filtered away is something that makes it clear that I'm looking at people's posts.''} (P8). However, three participants experienced usability challenges that the filtering was not accurate and did not sync up as experienced by three participants (P3, P4, P7). As P7 commented: \textit{``I would say it's not very accurate. It's not personal to me.''} In addition, the filtering caused challenges for tracking viewed posts because of the posts hiding and reordering, as what P8 experienced refreshing the page: \textit{``One that I was just on is now gone'' (P8)}. 

Participants (P2, P5, P8) suggested that the Filter Mode could have provided more options and combined multiple filters to achieve more customized results. For example, P2 compared the filter with Excel, and desired to have a wider range of filters such as \textit{``Quotes''} and \textit{``Meme}.''

\paragraph{Expert Feedback}
Filter Mode received positive feedback for its sorting feature and feed personalization (EP3--5) as well as its ease of use (EP3). As EP3 noted, \textit{``It’s just helpful to be able to narrow down, like, the type of content that you want, what’s most meaningful to them''} (EP3). However, some expert participants pointed out several usability challenges and provided suggestions for improvement, such as improving the accuracy of filtering results (EP2), making filter labels clearer (EP1, EP2), and increasing font size (EP3). 

\subsubsection{\textbf{Feedback for Facebook Customization}}
\paragraph{User Feedback} Seven participants (P1--3, P5--8) held positive views toward the aid, and one participant (P4) was neutral. Three participants (P3, P7, P8) appreciated the ability to control the newsfeed with the tool and mentioned that the aid could make them \textit{``more focused''} (P5) and hide information that they \textit{``don't really need''} (P3), such as advertisements. On the other hand, one participant (P4) found this customization was not necessary because their goal of using social media was to \textit{``check in and look around''} (P4), thus simplifying the information source would have a side effect for them to explore different content. P7 also wished that there were more options for the left-side filtering for him to \textit{``personalize it a little bit further''} (P7). P7 also desired more guidance on how to use the aid.

\paragraph{Expert Feedback}
Similar to the benefits mentioned by users with TBI, experts (EP1, EP3--5) noted that the tool simplified the newsfeed page and could help avoid distractions for users with TBI. EP1 noted, \textit{``For me as a user, I love this. I never looked at the stuff [extra fields on Newsfeed] before but I love it not being there. Look how much nicer that is to look at. I can ignore it with my own mind but I love not seeing it.''} Similarly, EP3 commented, \textit{``They have control over what they're seeing. This is very helpful''} (EP3). 
 However, experts (EP1, 2) mentioned the current design could be further improved by clarifying the wording of the options (EP2) and using icons to support the user's understanding of the control (EP1).

\section{Discussion}
 Overall, our findings reported feedback for our accessibility toolkit. Usability considerations reported by experts and users echoed the eight objectives from the W3C cognitive guidelines \cite{world2022all}, focusing on ease of use, prevention of errors, design consistency, clarity of information and personalization of the social media platform. Our findings highlighted the needs of building accessibility features in technologies to support cognitive and communication challenges for people with TBI. Communication support features in our system, such as sentiment and toxicity detection, writing support and post interpretation, can be applied to other types of online activities, such as sending instant messages, group communications in online communities, and reading online articles. Cognitive support features can be applied to other online platforms to simplify the website layouts, highlighting the needed features and providing specific step-by-step instructions to accomplish the tasks for individuals with TBI who are facing cognitive challenges. Our findings provided actionable steps for us to implement usability improvements and further iterate the system in preparation of extensive field testing. 

\subsection{Design Implications}
Our toolkit focused on providing communication support and cognitive support in social media use for individuals with TBI. Based on our findings, we drew the following five design implications for designing accessible social media: ensuring psychological safety, privacy control and protection, trade off between the business profits and user experience, mixed feedback in AI-tools, tool adoption for diverse TBI needs.

\paragraph{\textbf{Ensure Psychological Safety}}
Individuals with TBI can experience a variety of psychological challenges such as mood swings, depression, anxiety and agitation \cite{howlett2022mental}. Our user participants (P1, P2, P3) reported discomfort in seeing arguments or violent content on social media. Our expert participants (EP3, EP5) emphasized the importance of protecting TBI users' psychological safety and that the aid should give control to the content they are seeing.  Our toolkit employed the use of the ``\textit{positive}'' filter to avoid disturbing content in the newsfeed, and future design could further explore design solutions to increase the accessibility of content browsing and social communication for populations with psychological challenges. For example, the aid could predict the psychological safety level of a page that the user is going to visit for the user so they can decide if they want to continue or not. The aid can also provide support for the user to disclose their negative feelings if the user sees the content causing emotional swings. One of our TBI user participants (P2) mentioned that she would talk to her partner if she experienced emotional swing after seeing discomforting content on social media. Similarly, the aid could create a virtual character for the user to chat to and disclose their negative feelings if they do not have other people to talk to in the moment.

\paragraph{\textbf{Privacy Control and Protection}}
Expert participants cared for safety in using social media platforms and reported the possible oversharing activities by individuals with TBI in private messages and social media posts. This can be difficult to detect by individuals with TBI because they can be unaware of their behaviors or risks of oversharing on social media \cite{brunner2021rehabilitation}, especially for children and adolescents with brain injury. With more functional features provided on social media such as \textit{Market Place}, it is easy for users to talk to strangers, which can be risky for TBI users. To prevent oversharing social communications, the user needs to always pay attention to who they are talking to and what they are sharing with. First, the aid can alert the user if they are talking to strangers and ask for confirmation before they message out any personal or sensitive information. Second, the aid could help the user monitor if there are any potential risks in the ongoing conversation and provide resources and clear instructions on how to handle dangerous situations. Social communication can happen in a variety of forms on social media such as private messaging, group chat, post comments and sharing. Therefore, privacy protection should be supported across social media features.

\paragraph{\textbf{Trade-offs between Business Profits and Accessibility Needs}}
Advertisements  on social media were reported as the main challenge by our participants with TBI (six out of eight). Advertisements in the news feed can add to TBI users' cognitive load, cause difficulties for them to main focus, and have them lost in the navigation. Moreover, individuals with TBI often have short-term or long-term memory loss and can forget people in their friend list, thus causing challenges for the users to differentiate advertisements and posts from their social circle. It is acknowledged that advertisements are important for the company to make profits, however, they are also the major barriers for individuals with TBI to use social media platforms. Social media platforms could change the layout of the webpage to separate advertisements from the posts from the user's social circle. As suggested by our participant (P8), there could be a specific area on the webpage dedicated to advertisements to reduce distraction. In addition, social media platforms could offer different modes of advertisements and allow users to choose the one that meets their needs. One mode could be the existing approach that advertisements are integrated into the newsfeed. An alternative mode could be showing advertisements at certain times such as when opening a new webpage, or always displaying advertisements in a centralized area to make the social media site more accessible for users with cognitive impairments.

\paragraph{\textbf{Mixed Feedback for AI-based Tools for People with TBI}}
One interesting finding was participants' perceptions of using AI tools for communication support, in particular the use of sentiment analysis and toxicity detection in our toolkit. We demonstrated the potential of using these tools in supporting message constructions and interpretations during social communications for individuals with TBI. Nevertheless, we acknowledged the potential risks of adopting AI tools for people with TBI. The major concern expressed by our participants is the inaccurate results from the interpretation. We employed an off-the-shelf commercial AI tool for the analysis, however, an accuracy level of 80\% in the analysis caused much confusion for our participants as observed in the study. The inaccurate or biased results \cite{venkit2023automated} in AI-based tools can be risky for people with TBI because they can have difficulty understanding underlying messages and differentiating the biased views from the AI tool, which can shape the way they think in the long term. In addition, AI tools are becoming more pervasive and available to the public and an increasing number of rehabilitation features are built on top of them \cite{alsobhi2022facilitators}. The pervasiveness and overconfidence of AI tools can reduce people's initiatives and willingness to seek professional rehabilitation support provided by TBI experts and negatively affect their recovery process. 

\paragraph{\textbf{Tool Adoption with Diverse TBI Needs}}
Individuals with TBI can have diverse accessibility needs for using social media platforms. Through our study, participants with TBI reported specific desired features of the aids based on their personal needs, such as personalizing the Filter Mode with a ``Meme'' filter and using the Writing Aid to learn grammar (P2). Some participants preferred simplification of the interface using our cognitive support aids, while some preferred more complicated page layouts and richer interactions. Due to sensory challenges, participants have different preferences for the UI elements such as font and picture size. 

The accessibility toolkit design should provide customized options for the user to choose their desired features. Moreover, the toolkit should provide adaptive learning support for TBI users for better tool adaption. The learning support should provide clear step-by-step guidance for the use and should be always available when needed by the user due to the memory challenges commonly faced by people with TBI.

\subsection{Limitations \& Future Work}
Our work has several limitations in the capabilities of the aids as well as the conclusions we draw from our data. 

First, our aids relied on third-party APIs for text and sentiment analysis, which occasionally provided inaccurate interpretations. Significant errors in interpretation can impair user trust. Users with TBI might be particularly prone to the negative effects of errors due to their social communication challenges. \added{Moreover, \citet{hutchinson2020bias} revealed that \textit{Perspective API} used in the toolkit exhibited social biases towards disability-related terms and associated these terms with toxicity and negativity. Future work should incorporate bias mitigation techniques (\textit{e.g.}, \citet{cheng2022bias}) into the toolkit to prevent the harms from the socially biased analysis.}

Second, our aids are still proof-of-concept prototypes \added{and can be improved through integrating more advanced techniques. For example, Interpretation Aid can utilize tools for alt-text auto-generation \cite{wu2017alt, das2024alt} to improve its image description. In addition, usability issues of our prototypes can limit the effectiveness of the aids and have a negative effect on user perceptions.}  

Third, our evaluation took place in a laboratory environment for a short period of time, which offers little insight into long-term use and adoption patterns. \added{In our future work, we plan to conduct an in-the-wild study with an extended period of time to reflect on user's actual use of the toolkit in their social media platforms.}

Finally, due to the difficulty of recruiting users with TBI as well as TBI rehabilitation experts, our study included a relatively small sample size, which prevented meaningful quantitative data analysis. \added{Given the limited number of participants in our user evaluation ($N=8$), our primary goal is to understand the potential and initial user experiences through a qualitative approach. Data from our quantitative measures (\S\ref{sec-user-eval}) did not offer significant insights due to the potential lack of representativeness. However, the rich qualitative findings will guide the refinement of the toolkit for a more comprehensive and longer-term field study, where quantitative measures can be incorporated to assess the toolkit's effectiveness over an extended period.}

\section{Conclusion}
In this paper, we present the \textit{Social Media Accessibility and Rehabilitation Toolkit (SMART-TBI)}, which we built to support social communication and newsfeed browsing of social media users with TBI. The toolkit includes two communication support aids and three cognitive support aids. We evaluated the toolkit with eight social media users with TBI and five TBI experts. Our findings highlighted the effective features in the toolkit and pointed to areas of improvement for each aid. Based on our findings, we generated design implications to improve the accessibility of social media use by people with TBI, considering the psychological and privacy safety, the trade-offs between business profits and accessibility needs, mixed feedback for AI-based tools and tool adoption with diverse TBI needs. \added{These design implications can inform the building of accessible social media platforms for users with cognitive and communication difficulties.}

\begin{acks}
This work was funded by the National Institutes of Health (NIH R01-HD071089-06A1). Figure 1 modified an image by freepik for its design. We would like to thank our participants for their time and participation in this research study.
\end{acks}

\balance
\bibliographystyle{ACM-Reference-Format}
\bibliography{sample-base}

\appendix

\newpage
\section{Social Media Tasks}
\subsection{Writing Task} \label{apd: writing task}
Below are the three writing tasks that the participant completed with and without using the Writing Aid. 
\begin{enumerate}
    \item If you are going to share a movie on Facebook, how would you write about it? 
    \item If you are going to recommend a new restaurant on Facebook, how would you write about it?
    \item If you are going to write a post about the city you are living on Facebook, how would you write about it?
\end{enumerate}

Questions asked to the participant after they wrote each post:

\begin{enumerate}
    \item What emotions do you think of in the post? (open-ended)
    \item How negative or how positive do you think of this post? (open-ended)
    \item Please choose from Very Negative, Negative, Neutral, Positive and Very Positive.
\end{enumerate}

\subsection{Interpretation Task} \label{apd: interpretation task}

The following are the three posts that participants were asked to read and interpret with and without using the Interpretation Aid. 
\begin{enumerate}
\item Steamy day, but at least we have nature near where I work. Took a walk near the lakeshore and saw some muskrats swimming in the water. Took some photos and then went back to my office. My daily routine of living by the lake.

\item I can't believe squirrels ate the electrical wires in my car. Toyota coats the wires with a soy product to be "eco-friendly", and the squirrels ate it. It's going to cost me \$10,000 to replace. Squirrels are just rats with bushy tails and the city should ban them!

\item The week is half over and I still have so much to do! I attended a few meetings and then worked by the water the rest of the day. It is a beautiful day out today and working near nature helps me focus on everything I need to complete.
\end{enumerate}

Questions asked to the participant to interpretation each post:

\begin{enumerate}
    \item What emotions do you think of in the post? (open-ended)
    \item How negative or how positive do you think of this post? (open-ended)
    \item Please choose from Very Negative, Negative, Neutral, Positive and Very Positive.
\end{enumerate}

\section{Questionnaire}
\subsection{TBI Background}
What was the main cause of your brain injury?
\begin{itemize}
    \item Motor vehicle crashes involving occupants or pedestrians
    \item Sports and recreation injuries (e.g. sports concussions, bicycling injuries)
    \item Assaults and violence (e.g. domestic violence, abuse, gunshot wounds/firearm injuries)
    \item Shaken Baby Syndrome- Abusive Head Trauma (AHT) or inflicted Traumatic Brain Injury (iTBI)
    \item Blunt trauma- struck by or against an object
    \item Penetrating or open head wounds (e.g. lacerations)
    \item Explosive blasts (e.g. Improvised Explosive Devices)
\end{itemize}

What kinds of challenges have you experienced after you acquired a brain injury? (You can select multiple answers)
\begin{itemize}
    \item Short-term or long-term memory loss
    \item Impaired judgment and perception
    \item Trouble concentrating or paying attention
    \item Difficulty with language or speech production and thought processing
    \item Spatial disorientation
    \item Difficulty organizing or problem solving
    \item Sensory loss or impairment (vision, hearing, etc.)
    \item Headaches or migraines
    \item Decreased motor abilities
    \item Depression
    \item Anxiety, restlessness, agitation, frustration, impatience
    \item Lack of motivation
    \item Reduced level of self-esteem
    \item Mood swings
    \item Impulsiveness and lack of inhibition
    \item Personality changes
    \item Emotional flatness and passivity
    \item Other
    \item Prefer not to answer
\end{itemize}

\section{User Evaluation Result}

\subsection{SUS score} \label{apd: SUS scores}
\begin{table*}[]
    \caption{TBI user participants' SUS feedback on each aid. D: \textit{Disagree}, N: \textit{Neutral}, A: \textit{Agree}. Data represent the percentage of users choosing the option. E.g., For the statement ``I think that I would like to use this system frequently,'' 0.25 under D in Writing aid indicates that 25\% of participants chose Disagree for this statement; 0.25 under N indicates that 25\% of participants chose Neutral for this statement; and 0.5 under A indicates that 50\% of participants chose Agree for this statement.}
    \label{tab: user SUS result}
    \centering
    \renewcommand{\arraystretch}{1.2}
    \footnotesize
    \begin{tabular}{p{0.2\linewidth}p{0.02\linewidth}p{0.02\linewidth}p{0.02\linewidth}p{0.02\linewidth}p{0.02\linewidth}p{0.02\linewidth}p{0.02\linewidth}p{0.02\linewidth}p{0.02\linewidth}p{0.02\linewidth}p{0.02\linewidth}p{0.02\linewidth}p{0.02\linewidth}p{0.02\linewidth}p{0.02\linewidth}}
\toprule
SUS result                                                                                                        & \multicolumn{3}{c}{\textbf{Writing Aid}}                                                        & \multicolumn{3}{c}{\textbf{Interpretation Aid}}                                                 & \multicolumn{3}{c}{\textbf{Focus Mode}}                                                         & \multicolumn{3}{c}{\textbf{Filter Mode}}                                                        & \multicolumn{3}{c}{\textbf{Facebook Customization}}                                                   \\
Percentage of participants choosing \textit{Disagree}, \textit{Neutral}, and \textit{Agree}                                                                                                         & \multicolumn{1}{c}{D} & \multicolumn{1}{c}{N} & \multicolumn{1}{c}{A} & \multicolumn{1}{c}{D} & \multicolumn{1}{c}{N} & \multicolumn{1}{c}{A} & \multicolumn{1}{c}{D} & \multicolumn{1}{c}{N} & \multicolumn{1}{c}{A} & \multicolumn{1}{c}{D} & \multicolumn{1}{c}{N} & \multicolumn{1}{c}{A} & \multicolumn{1}{c}{D} & \multicolumn{1}{c}{N} & \multicolumn{1}{c}{A} \\
\toprule
I think that I would like to use this system frequently.                                  & 0.25                         & 0.25                        & 0.50                      & 0.75                         & 0.00                        & 0.25                      & 0.75                         & 0.00                        & 0.25                      & 0.63                         & 0.13                        & 0.25                      & 0.17                         & 0.00                        & 0.83                      \\
\midrule
I found the system unnecessarily complex.                                              & 0.88                         & 0.13                        & 0.00                      & 0.88                         & 0.13                        & 0.00                      & 0.50                         & 0.25                        & 0.25                      & 0.88                         & 0.00                        & 0.13                      & 0.83                         & 0.00                        & 0.17                      \\
\midrule
I thought the system was easy to use.                                                    & 0.13                         & 0.13                        & 0.75                      & 0.00                         & 0.00                        & 1.00                      & 0.00                         & 0.25                        & 0.75                      & 0.00                         & 0.13                        & 0.88                      & 0.00                         & 0.00                        & 1.00                      \\
\midrule
I think that I would need the support of a technical person to be able to use this system. & 1.00                         & 0.00                        & 0.00                      & 1.00                         & 0.00                        & 0.00                      & 0.88                         & 0.00                        & 0.13                      & 1.00                         & 0.00                        & 0.00                      & 1.00                         & 0.00                        & 0.00                      \\
\midrule
I found the various functions in this system were well integrated.                         & 0.00                         & 0.13                        & 0.88                      & 0.00                         & 0.13                        & 0.88                      & 0.25                         & 0.25                        & 0.50                      & 0.25                         & 0.00                        & 0.75                      & 0.33                         & 0.00                        & 0.67                      \\
\midrule
I thought there was too much inconsistency in this system.                                & 1.00                         & 0.00                        & 0.00                      & 0.75                         & 0.00                        & 0.25                      & 0.50                         & 0.00                        & 0.50                      & 0.50                         & 0.00                        & 0.50                      & 0.83                         & 0.00                        & 0.17                      \\
\midrule
I would imagine that most people would learn to use this system very quickly.           & 0.00                         & 0.00                        & 1.00                      & 0.00                         & 0.00                        & 1.00                      & 0.25                         & 0.13                        & 0.63                      & 0.00                         & 0.13                        & 0.88                      & 0.00                         & 0.00                        & 1.00                      \\
\midrule
I found the system very cumbersome to use.                                                & 0.63                         & 0.00                        & 0.38                      & 0.75                         & 0.00                        & 0.25                      & 0.50                         & 0.25                        & 0.25                      & 0.63                         & 0.00                        & 0.38                      & 0.67                         & 0.00                        & 0.33                      \\
\midrule
I felt very confident using the system.                                                   & 0.00                         & 0.00                        & 1.00                      & 0.13                         & 0.00                        & 0.88                      & 0.38                         & 0.00                        & 0.63                      & 0.13                         & 0.13                        & 0.75                      & 0.00                         & 0.00                        & 1.00                      \\
\midrule
I needed to learn a lot of things before I could get going with this system.             & 1.00                         & 0.00                        & 0.00                      & 0.88                         & 0.00                        & 0.13                      & 0.75                         & 0.00                        & 0.25                      & 0.88                         & 0.00                        & 0.13                      & 0.83                         & 0.00                        & 0.17              \\
\bottomrule
\end{tabular}
\end{table*}

\end{document}